\documentclass[showpacs,preprintnumbers,
superscriptaddress]{revtex4}

\usepackage[dvips]{graphics}

\newcommand{\AP}[3]{Ann.\ Phys.\ {\bf #1},\ #2 (#3)}
\newcommand{\NPA}[3]{Nucl.\ Phys.\ {\bf A#1},\ #2 (#3)}
\newcommand{\NPB}[3]{Nucl.\ Phys.\ {\bf B#1},\ #2 (#3)}

\newcommand{\PLB}[3]{Phys.\ Lett.\ B\ {\bf #1},\ #2 (#3)}
\newcommand{\PR}[3]{Phys.\ Rep.\ {\bf #1},\ #2 (#3)}
\newcommand{\PRL}[3]{Phys.\ Rev.\ Lett.\ {\bf #1},\ #2 (#3)}

\newcommand{\PRD}[3]{Phys.\ Rev.\ D\ {\bf #1},\ #2 (#3)}
\newcommand{\JPG}[3]{J.\ Phys.\ G\ {\bf #1},\ #2 (#3)}
\newcommand{\RPP}[3]{Rep.\ Prog.\ Phys.\ {\bf #1},\ #2 (#3)}
\newcommand{\ZPC}[3]{Z.\ Phys.\ C\ {\bf #1},\ #2 (#3)}
\newcommand{\EPJC}[3]{Eur.\ Phys.\ J.\ C\ {\bf #1},\ #2 (#3)}
\newcommand{\PPNP}[3]
{Prog.\ Part.\ Nucl.\ Phys.\ {\textbf #1},\ #2 (#3)}
\newcommand{\APPB}[3]{Acta.\ Phys.\ Polon.\ B\ {\bf #1},\ #2 (#3)}

\newcommand{\ibid}[3]{{\bf #1},\ #2 (#3)}

\renewcommand\a{\alpha}
\renewcommand\b{\beta}

\renewcommand\d{\delta}

\newcommand\m{\mu}
\newcommand\n{\nu}

\newcommand\ra{\rightarrow}

\newcommand{\ovl}[1]{\overline{#1}}

\newcommand{\non}{\nonumber\\}

\newcommand{\bee}{\begin{equation}}
\newcommand{\eee}{\end{equation}}
\newcommand{\bea}{\begin{eqnarray}}
\newcommand{\eea}{\end{eqnarray}}
\newcommand{\ba}[1]{\begin{array}{#1}}
\newcommand{\ea}{\end{array}}

\newcommand{\eqrf}[1]{Eq.\ (\ref{#1})}
\newcommand{\eqrfm}[2]{Eqs.\ (\ref{#1}-\ref{#2})}
\newcommand{\eqrftw}[2]{Eqs.\ (\ref{#1}) and (\ref{#2})}

\begin{document}

\preprint{CERN-TH/2002-015}

\title{From the Dyson-Schwinger to the Transport Equation
\\in the Background Field Gauge of QCD }

\author{Qun Wang}
\email{qwang@th.physik.uni-frankfurt.de}
\affiliation{
Institute f\"ur Theoretische Physik, J. W.
Goethe-Universit\"at, D-60054 Frankfurt, Germany
}
\affiliation{
Physics Department, Shandong University,
Jinan, Shandong 250100, People's Republic of China
}
\author{Krzysztof Redlich}
\email{Krzysztof.Redlich@cern.ch}
\affiliation{
Theory Division, CERN, CH-1211 Geneva 23, Switzerland
}
\affiliation{
Institute for Theoretical Physics, University of Wroclaw,
PL-50204  Wroclaw, Poland
}
\author{Horst St\"ocker}
\email{stoecker@th.physik.uni-frankfurt.de}
\author{Walter Greiner}
\email{greiner@th.physik.uni-frankfurt.de}
\affiliation{
Institute f\"ur Theoretische Physik,
J. W. Goethe-Universit\"at, D-60054 Frankfurt, Germany
}

\date{\today}

\begin{abstract}

The non-equilibrium quantum field dynamics is usually described in
the closed-time-path formalism. The initial state correlations are
introduced into the generating functional by non-local source
terms. We propose a functional approach to the Dyson-Schwinger
equation, which treats the non-local and local source terms in the
same way. In this approach, the generating functional is
formulated for the connected Green functions and
one-particle-irreducible vertices. The great advantages of our
approach over the widely used two-particle-irreducible
method are that it is much simpler
and that it is easy to implement the procedure in a computer
program to automatically generate the Feynman diagrams for a given
process. The method is then applied to a pure gluon plasma to
derive the gauge-covariant transport equation from the
Dyson-Schwinger equation in the background covariant gauge. We
discuss the structure of the kinetic equation
and show its relationship with the classical one.
We derive the gauge-covariant collision part and
present an approximation in the vicinity of equilibrium.
The role of the non-local source kernel in the
non-equilibrium system is discussed
in the context of a free scalar field.

\end{abstract}


\maketitle

\section{Introduction}

The ultimate goal of ultra-relativistic nucleus-nucleus collisions
is to study the properties of strongly interacting matter under
extreme conditions of high energy density \cite{rev}. Quantum
Chromodynamics (QCD) predicts that strongly interacting matter
undergoes a phase transition from a state of hadronic constituents
to a plasma of unbounded quarks and gluons (QGP) \cite{karsch}.
The QGP is considered as a partonic system being at (or close to)
local thermal equilibrium. Thus, to study the conditions for the
possible formation of a QGP in heavy-ion collisions one needs to
address the question of thermalization of the initially produced
partonic medium \cite{stock,baier}. There
are various theoretical predictions suggesting that the
appearance of the QGP could modify the properties of physical
quantities that are measured in heavy-ion collisions. The
suppression of charmonium production \cite{hs} was argued to be a
possible consequence of the collective effects in the thermalized
and deconfined medium. The jet quenching \cite{mgxw,baier1} was
predicted to be due to the radiative energy loss of partons
penetrating the QGP. Strangeness \cite{raf} and
dilepton \cite{vesa} production yields
could be modified in the thermal QCD medium.
A variety of models for the initial conditions in ultra-relativistic
heavy-ion collisions suggest that at the early stage the medium is
dominated by the gluon degree of freedom \cite{sat}. The transport
equation for a pure gluon plasma is thus of special interest.

The usual treatment of the gluon transport equation is based on
the decomposition of the gluon field into a mean field and a
quantum fluctuation. Under this
approximation the gluon transport equation then describes the
kinetics of the quanta in the classical mean field
\cite{heinz,geiger,elze88}.
This picture is somewhat similar to what was used
while studying the energy loss of the fast parton moving in
the soft mean field \cite{mgxw,baier1}.
To include the classical chromofield into QCD in
a proper way, one uses the background field method of QCD
(BG-QCD) introduced by DeWitt and 't Hooft
\cite{dewitt,thooft,abb81,soh86,zub75}. The advantage of BG-QCD is
that it is formulated in an explicit gauge-invariant manner.
The BG-QCD is a very suitable method to describe the properties of
a QCD medium created in the initial state of heavy-ion collisions.
The time evolution of a quantum system being off equilibrium
can be, in principle, obtained by solving
the Dyson-Schwinger equation (DSE) defined on a closed-time-path
(or the Kadanoff-Baym equations). If the
kinetic scale, describing long-range correlations, is much larger
than the scale of quantum fluctuations, the DSE may be reduced
into a much simpler form of the transport equation by a gradient
expansion \cite{elze86,ctp,calzetta,mao}. 
To derive the transport equation in
the presence of a classical chromofield in an explicit 
gauge-invariant or covariant way, one thus combines the BG-QCD method
and the closed-time-path (CTP) formalism.

To our knowledge, the first study of the kinetics of a classical
particle with non-Abelian gauge degrees of freedom propagating in
a non-Abelian classical gauge field is due to Wong \cite{wong}.
There were many efforts in the literature to derive the transport
equation for the QCD medium \cite{jalilian,kelly,litim}, 
particularly by using the
BG-QCD method \cite{elze90,blaizot,blaizot1}.
To our knowledge, the first application of
this method was done by Elze \cite{elze90}
to derive the transport equation for gluons.
His approach was based on the Yang-Mills field equation and the
second quantization in the operator representation
of the quantum field theory. However
the equation obtained is in a complicated form.
In \cite{geiger}, the transport equation for
the gluon was derived in the conventional
QCD by using the light-cone gauge and CTP formalism.
The gluon field was decomposed into a hard and a soft part 
treated as the classical field. This decomposition,
however, was not done in a gauge-covariant way.
Most recently Blaizot and Iancu have derived
the Boltzmann equation for the QCD plasma \cite{blaizot}, 
applying the CTP formalism and the BG-QCD
method which guarantees the gauge-covariant
decomposition of the gauge field.

To derive the Boltzmann equation one needs, in general, to make a
set of approximations. This usually involves a gradient expansion
of the DSE and the perturbative derivation of the collision term.
In Ref.~\cite{blaizot} an additional assumption has been made that
the system is close to equilibrium. Consequently, the
transport equation for the QCD plasma obtained in
Ref.~\cite{blaizot} was linearized with respect to the
off-equilibrium fluctuations.

In this paper we propose a different approach to derive the
Boltzmann equation for the gluon plasma in the BG-QCD. First we
apply the functional approach to the DSE that treats the non-local
and local source terms in the same way. We strictly stick to the
functional definition of the one-particle-irreducible (1-PI)
vertices and the connected Green functions (CGF). Furthermore, we
use the DeWitt notation, which, in our opinion, results in a simple
structure for the generating functionals. The current approach has
a great advantage over the widely used two-particle-irreducible
method that it is much simpler and can automatically generate all
necessary Feynman diagrams.

In a heuristic discussion on the role
of the non-local kernel for a free scalar field, we show that
if the initial time is in the remote
past the kernel provides only a correction to the homogeneous
solution of the DSE and preserves its structure. We also see in
this simple model that if the initial time is not in the
remote past, the non-local kernel brings the time dependence to
the DSE and  breaks the assumed structure of the homogeneous
solution.

From the DSE, we derive the transport equation in a gauge-covariant
way. Our derivation is quite general as it does not require any
additional assumptions, such as a special form for the
gauge-covariant Green function (GF) or that the system is 
near equilibrium. Consequently our equation 
is not linearized with respect to the off-equilibrium
fluctuations. We use the background {\em covariant} instead of the
background {\em Coulomb} gauge as was used
in Ref.\ \cite{blaizot}. Therefore our
results preserve an explicit Lorentz covariance and have a compact
structure. However, we have to include the ghost fields to cancel
the non-physical degrees of freedom of the gauge field.
We note that the resulting kinetic equation
has a structure similar to the one previously
derived in \cite{mrow}, based on \cite{elze88} in QCD and
assuming the 2-point gauge-covariant GF (or the Wigner function)
to be proportional to the quadratic product of the generators
for the fundamental color representation.

We discuss the structure of the kinetic part of
the transport equation derived here and compare it with
the classical kinetic equation. In the quantum case the
kinetic equation describes the time evolution of the gauge-covariant
Wigner function, which is a matrix in the adjoint color space.
Therefore, it contains many non-Abelian features, which are absent
from the well known classical equation. However, a notable result
is that, as in the classical case, it contains a term that
corresponds to the color precession \cite{old}.
This is the non-Abelian analogue to the Larmor precession
for particles with magnetic moments in an external magnetic field.
We argue that this term is necessary to preserve
the gauge covariance of the resulting
transport equation. We also discuss the structure of the kinetic
part if the system exhibits only a small deviation from 
equilibrium.

We derive the gauge-covariant collision part of the Boltzmann
equation and present its linearized form with respect to the
off-equilibrium fluctuations.
Finally applying the transversality condition for 
the gauge-covariant GF, together with some other
approximations, the collision part is further simplified and shows
the explicit collision and damping terms.

The paper is organized as follows. In Section II we introduce the
basic concept of BG-QCD. In Section III we present two equivalent,
in a path-integral sense, methods to derive the classical
equation of motion for the gluon.
In Section IV, we describe the functional approach to the DSE in
the vacuum. We derive the DSE for the 2- and 3-point GF
in the background covariant gauge. In Section V we
present a derivation of the DSE for the non-equilibrium gluon
plasma within the CTP formalism. Section VI contains a heuristic
discussion on the role of the non-local source kernel. In Section
VII we derive the transport equation from the DSE, applying the
gradient expansion. In Section VIII the gauge-covariant
collision part is derived and its structure under some
approximations is discussed.

In the paper we use $g_{\m\n}={\rm diag}(1,-1,-1,-1)$ as the
metric tensor. The Lorentz indices are written as subscripts and
color indices as superscripts to the relevant quantities.
For a pure gluon plasma the color field transforms only under the
$SU(3)_c$ adjoint representation, thus all color indices are
adjoint ones. We use the following notation for the gauge field
and its field strength tensor: 
$A_{\m}\equiv A_{\m}^aT^a$ and $F_{\m\n}[A]\equiv 
F_{\m\n}^a[A]T^a$, where $(T^a)^{ij}=if^{iaj}$ are the generators
of the $SU(3)_c$ adjoint representation and $f^{abc}$ are the
$SU(3)_c$ structure constants. The two-point GF or self-energy
(SE) are treated as matrices, thus the color and/or Lorentz
indices are sometimes omitted.

For convenience, we list all abbreviations that we use in
this paper: background gauge QCD (BG-QCD), Dyson-Schwinger
equation (DSE), closed-time-path (CTP), Green function (GF),
connected Green function (CGF), one-particle-irreducible (1-PI)
and self-energy (SE).

\section{Background Field Method}

Any physical quantity calculated in QCD is gauge
invariant and independent of the particular gauge chosen. The
classical Lagrangian for the gauge field exhibits an explicit gauge
invariance. However, when quantizing the theory in a particular
gauge, one introduces the gauge fixing and ghost terms
that break the explicit gauge invariance of the Lagrangian. The
background field method is such a technique that allows 
a gauge (background gauge) to be fixed 
without losing the classical gauge invariance.

In BG-QCD the conventional gluon field is expressed as a sum
of a classical background field $A$ and a quantum fluctuation
$Q$. The action is given by \cite{abb81,soh86,zub75}
\bea
&&S=S_0+S_{fix}+S_{ghost}+S_{src} ,\nonumber\\
&&S_0=-\frac 14 \int d^4x (F_{\mu \nu}^i[A]+D_{\mu}^{ij}[A]Q_{\nu}^j-
D_{\nu}^{ij}[A]Q_{\mu}^j+gf^{ijk}Q_{\mu}^jQ_{\nu}^k)^2 ,
\nonumber\\
&&S_{fix}=-\frac {1}{2\alpha}\int d^4x
(D_{\mu}^{ij}[A]Q^{\mu ,j})^2 ,\nonumber\\
&&S_{ghost}=\int d^4x \overline{C}^i D_{\mu}^{ij}[A]
D^{\mu ,jk}[A+Q]C^k ,\nonumber\\
&&S_{src}=\int d^4x (J_{\mu}^i Q^{\mu ,i} +\overline{\xi}^i C^i
+\overline{C}^i \xi ^i) \;.
\label{s}
\eea
The generating functional for the GF reads
\begin{equation}
\label{Z} Z[A,J,\xi ,\overline{\xi}]=\int [dQ][dC][d\overline{C}]
\exp (iS),
\end{equation}
where the quantum fluctuations $Q_{\mu}^i$ of the gluon field are the
integration variables in the functional integral;
$C^i, \overline{C}^i$ are the ghost and antighost fields;
$\overline{\xi}^i, \xi ^i$
are the external sources coupling to the ghost and
antighost fields respectively.
The covariant derivative $D_{\mu}^{ij}[A]$ is
defined by $D_{\mu}^{ij}[A]\equiv
\partial _{\mu}\delta ^{ij}-ig(T^{a})^{ij}A_{\mu}^a$.
The field strength tensor $ F_{\mu \nu}^i[A]$ for the
background field is given by
$F_{\mu \nu}^i[A]=\partial _{\mu}A_{\nu}^i
-\partial _{\nu}A_{\mu}^i+gf^{ijk}A_{\mu}^jA_{\nu}^k$.

There are two types of gauge transformations
that leave $Z$ invariant \cite{zub75}.
The type I transformations are given by
\bea
A'_{\mu}&=&UA_{\mu}U^{-1}+\frac ig U\partial _{\mu}U^{-1},
\nonumber\\
Q'_{\mu}&=&UQ_{\mu}U^{-1},\,\,\, J'_{\mu}=UJ_{\mu}U^{-1},
\nonumber\\
C'_{\mu}&=&UCU^{-1},\,\,\, \overline{\xi}'=U\overline{\xi}U^{-1},
\nonumber\\
\overline{C}'&=&U\overline{C}U^{-1},\,\,\, \xi '=U\xi U^{-1}\;,
\label{t1}
\eea
where $U(x)=\exp (ig\omega ^a(x)T^a)$.
For an infinitesimal gauge transformation
the fields transform as
\bea
&&\delta A_{\mu}^i=D_{\mu}^{ij}[A]\omega ^j\;,\nonumber\\
&&\delta Q_{\mu}^i=gf^{ijk}Q_{\mu}^j \omega ^k ,\,\,\,\,
\delta J_{\mu}^i=gf^{ijk}J_{\mu}^j \omega ^k \;,\nonumber\\
&&\delta C^i=gf^{ijk}C^j \omega ^k, \,\,\,\,
\delta \overline{\xi}^i=gf^{ijk}\overline{\xi}^j \omega ^k \;,
\nonumber\\
&&\delta \overline{C}^i=gf^{ijk}\overline{C}^j \omega ^k,\,\,\,
\delta \xi ^i=gf^{ijk}\xi ^j \omega ^k \;.
\label{s-tr}
\eea
In the type II transformations,
the background field does not change,
but the gluon field transforms as
$Q_{\mu}'=U(A_{\mu}+Q_{\mu})U^{-1}
+ig^{-1}U\partial _{\mu}U^{-1}$.
In this paper, as in Ref.\ \cite{elze90},
type I transformations are relevant
where the background field transforms
in a conventional way and the gluon
transforms like a matter field.

In BG-QCD, we define
the generating functional $W[A,J,\xi,\overline{\xi}]$ for the CGF
and $\Gamma [A,\langle Q\rangle ,\langle C\rangle , \langle
\overline{C}\rangle ]$ for the 1-PI vertex.
Both $W$ and $\Gamma $ are gauge-invariant functionals, thus
should be invariant under the gauge transformations
(\ref{t1}) and (\ref{s-tr}). Consequently, one can formulate 
for $\Gamma$/$W$ a Ward identity that
corresponds to this invariance \cite{zub75}.

\section{Equation of motion for gluon and classical field}

The equation of motion for the gluon field $Q$ is obtained by using
the conventional variation approach. Shifting the gluon field
$Q_{\mu}^i(x)$ by a small amount $v_{\mu}^i(x)$:
${Q^i}'_{\mu}(x)=Q_{\mu}^i(x)+v_{\mu}^i(x)$, the variation
of the functional $W$ is
\begin{equation}
\delta W=-i\delta \ln Z=Z^{-1}
\int [dQ][dC][d\overline{C}]
\int d^{4}x
\frac{\delta S}{\delta Q_{\mu}^i}v_{\mu}^i(x)\exp (iS)\;.
\end{equation}
Because we only shift a variable of integration
in the functional integral,
there should be no change for $W$.
This requirement leads to the equation of motion
for $Q_{\mu}^i$:
\bee
\bigg\langle\frac{\delta S}{\delta Q_{\mu}^i(x)}\bigg\rangle
\equiv Z^{-1} \int [dQ][dC][d\overline{C}]
\frac{\delta S}{\delta Q_{\mu}^i}\exp (iS)=0\;,
\eee
where the expectation value
$\langle\cdots\rangle$ is taken in the path
integral sense. In the absence of the external sources $J$,
$\xi$ and $\overline{\xi}$ the equation of motion becomes
\bee
\label{eqm}
D_{\nu}^{ij}[A]F^{\nu\mu,j}[A]
=\langle j^{\mu ,i}\rangle \;,
\eee
with the induced current $j$ is given by:
\bea
j&=&j_0+j_{gf}\;, \nonumber\\
j_0^{\mu ,i}&=&
-gf^{abc}D_{\nu}^{ia}[A](Q^{\nu, b}Q^{\mu ,c})
-gf^{ida}Q_{\nu}^d(D^{\nu ,ab}[A]Q^{\mu, b}
-D^{\mu ,ab}[A]Q^{\nu, b})\nonumber\\
&&-g^2f^{ida}f^{abc}Q_{\nu}^{d}Q^{\nu ,b}Q^{\mu ,c}
-D_{\nu}^{ij}[A](D^{\nu ,jk}[A]Q^{\mu, k}
-D^{\mu ,jk}[A]Q^{\nu, k})\nonumber\\
&&-gf^{ijk}Q_{\nu}^jF^{\nu\mu,k}[A]\;, \nonumber\\
j_{gf}^{\mu ,i}&=&gf^{kil}\overline{C}^j
\stackrel{\leftarrow}{D}\raisebox{0.5ex}{$^{\mu,kj}$}[A]C^l
-\frac{1}{\alpha}D^{\mu,ij}[A]D^{\nu,jk}[A]Q_{\nu}^k \;,
\label{jq}
\eea
where $j_0$ comes from $S_0$ and $j_{gf}$ from the
ghost and gauge-fixing terms. One can verify that $j$
transforms like a matter field.

In fact, we have $\langle\delta S/\delta Q\rangle =\langle\delta
S/\delta A\rangle =0$. The proof can be found in
Refs. \cite{zub75} and \cite{lee73}.
From the condition that $\langle\delta S/\delta A\rangle=0$,
and after setting all external sources to zero, we obtain
\begin{equation}
\label{eqm1}
D_{\nu}^{ij}[A]F^{\nu\mu,j}[A]
=\langle j'^{\mu ,i}\rangle \;,
\end{equation}
where $j'$ is defined by
\bea
j'&=&j_0+j'_{gf}\;,\nonumber\\
j'^{\mu ,i}_{gf}&=&-gf^{kil}\overline{C}^j
\stackrel{\leftarrow}{D}\raisebox{0.5ex}{$^{\mu,kj}$}[A]C^l
-gf^{ijk}\overline{C}^jD^{\mu,kl}[A+Q]C^l\nonumber\\
&&+\frac 1{\alpha}gf^{ijk}Q^{\mu,k}D_{\nu}^{jl}[A]Q^{\nu,l} \;.
\label{ja}
\eea
Note that $j'$ is different from $j$
because $j_{gf}\neq j'_{gf}$. However, the expectation
values of these two currents are equal. Following from the identity
$D_{\mu}^{ij}[A]D_{\nu}^{jk}[A]F^{\nu\mu,k}[A]=0$, we also have
$D_{\mu}^{ij}[A]\langle j^{\mu ,j}\rangle =D_{\mu}^{ij}[A]\langle
j'^{\mu ,j}\rangle=0$.

\section{A Functional Approach to the 
Dyson-Schwinger Equation in BG-QCD}

In the previous section, we have discussed the classical
equation of motion for the background and quantum fields.
In the following we will show how to obtain the DSE.
We will use a convenient functional approach, which has the
advantage that the non-local and the local terms are treated in the
same way. In our derivation we always use
the notation introduced by DeWitt \cite{dewitt1}, 
which gives each formula a simple face and a clear structure.
In DeWitt notation the classical action (\ref{s})
can be written in the following form:
\bea
S&=&S_{0}+S_{src}\;, \nonumber\\
S_{0}&=&\frac 12 \Gamma _{mn}^{(0)}(A^2)A_mA_n
+\frac 16 \Gamma _{mnp}^{(0)}(A^3)A_mA_nA_p
+\frac 1{24} \Gamma _{mnpq}^{(0)}(A^4)A_mA_nA_pA_q
\nonumber\\
&&+\frac 12 \Gamma _{mn}^{(0)}(Q^2)Q_mQ_n
+\frac 16 \Gamma _{mnp}^{(0)}(Q^3)Q_mQ_nQ_p
+\frac 1{24} \Gamma _{mnpq}^{(0)}(Q^4)Q_mQ_nQ_pQ_q
\nonumber\\
&&+\Gamma _{mn}^{(0)}(AQ)A_mQ_n
+\frac 12 \Gamma _{mnp}^{(0)}(AQ^2)A_mQ_nQ_p
+\frac 12 \Gamma _{mnp}^{(0)}(A^2Q)A_mA_nQ_p
\nonumber\\
&&+\frac 16 \Gamma _{mnpq}^{(0)}(A^3Q)A_mA_nA_pQ_q
+\frac 16 \Gamma _{mnpq}^{(0)}(AQ^3)A_mQ_nQ_pQ_q
\nonumber\\
&&+\frac 14 \Gamma _{mnpq}^{(0)}(A^2Q^2)A_mA_nQ_pQ_q
\nonumber\\
&&+\Gamma _{mn}^{(0)}(\ovl{C}C)\ovl{C}_mC_n
+\Gamma _{mnp}^{(0)}(\ovl{C}CA)\ovl{C}_mC_nA_p
+\Gamma _{mnp}^{(0)}(\ovl{C}CQ)\ovl{C}_mC_nQ_p
\nonumber\\
&&+\Gamma _{mnpq}^{(0)}(\ovl{C}CAQ)\ovl{C}_mC_nA_pQ_q
+\frac 12\Gamma _{mnpq}^{(0)}(\ovl{C}CQ^2)\ovl{C}_mC_nQ_pQ_q
\;, \nonumber\\
S_{src}&=&J_mQ_m+\ovl{\xi}_mC_m+\ovl{C}_m\xi _m \;,
\eea
where subscripts $m,n,p,q$ represent all necessary indices,
including colour, Lorentz and space-time coordinates.
The fractions in front
of some terms are symmetry factors; $\Gamma _{mnp}^{(0)}(AQ^2)$,
for example, is the bare vertex attaching one $A$ and two $Q$'s.
The explicit forms of these bare vertices are given in Appendix A.
We use the convention that the repetition of two indices stands
for a sum or an integral. Note that the definition of $S_{0}$ has
changed from Eq.\ (\ref{s}).

From Section III, one knows that the expectation value of the first
derivative of $S$ with respect to $Q$ leads to the classical
equation of motion. In DeWitt notation this derivative
is written as:
\bea
\frac{\delta S}{\delta Q_m}&=&
\Gamma _{mn'}^{(0)}(Q^2)Q_{n'}
+\Gamma _{n'm}^{(0)}(AQ)A_{n'}
+\Gamma _{m'n'm}^{(0)}(AQ^2)A_{m'}Q_{n'}\nonumber\\
&&+\frac 12 \Gamma _{m'n'm}^{(0)}(A^2Q)A_{m'}A_{n'}
+\frac 16 \Gamma _{m'n'p'm}^{(0)}(A^3Q)A_{m'}A_{n'}A_{p'}
+\frac 12 \Gamma _{m'n'p'm}^{(0)}(A^2Q^2)A_{m'}A_{n'}Q_{p'}
\nonumber\\
&&+\frac 12 \Gamma _{mn'p'}^{(0)}(Q^3)Q_{n'}Q_{p'}
+\frac 16 \Gamma _{mn'p'q'}^{(0)}(Q^4)Q_{n'}Q_{p'}Q_{q'}
+\frac 12 \Gamma _{m'n'p'm}^{(0)}(AQ^3)A_{m'}Q_{n'}Q_{p'}
\nonumber\\
&&+\Gamma _{m'n'm}^{(0)}(\ovl{C}CQ)\ovl{C}_{m'}C_{n'}
+\Gamma _{m'n'p'm}^{(0)}(\ovl{C}CAQ)\ovl{C}_{m'}C_{n'}A_{p'}
\nonumber\\
&&+\Gamma _{m'n'p'm}^{(0)}
(\ovl{C}CQ^2)\ovl{C}_{m'}C_{n'}Q_{p'}+J_m\;.
\label{ds-q}
\eea
The classical equation of motion then reads
\bee
\bigg\langle\frac{\delta S} {\delta Q_m}\bigg\rangle
=0\,\,\,{\rm or}\,\,\,
\bigg\langle\frac{\delta S_{0}} {\delta
Q_m}\bigg\rangle =-J_m \;.
\eee
The explicit form of the equation
of motion in the absence of $J_m$ is given by \eqrf{eqm}.
From the 1-PI generating functional,
we also have $\delta \Gamma /\delta
\langle Q_m\rangle =-J_m$, consequently we have
\bee
\label{g-s}
\bigg\langle\frac{\delta S_{0}}{\delta
Q_m}\bigg\rangle
=\frac{\delta \Gamma }{\delta \langle Q_m\rangle } \;.
\eee

From now on we will always refer to $S_{0}$ instead of $S$, 
we therefore drop the subscript and simply denote it as $S$.
Taking the derivative of \eqrf{g-s} with respect to
$\langle Q_p\rangle $, we have
\bea \frac{\delta ^2\Gamma} {\delta
\langle Q_p\rangle \delta \langle Q_m\rangle } &=&\frac{\delta
J_n}{\delta \langle Q_p\rangle } \frac{\delta }{\delta J_n}
\bigg\langle\frac{\delta S}{\delta Q_m}\bigg\rangle =-\frac{\delta
^2\Gamma} {\delta \langle Q_p\rangle \delta \langle Q_n\rangle }
\frac{\delta }{\delta J_n} \bigg\langle\frac{\delta S}{\delta
Q_m}\bigg\rangle
\nonumber\\
&=&-i\Gamma _{pn}(Q^2)\Big[ \langle I^{1}_mQ_n\rangle -\langle
Q_n\rangle \langle I^{1}_m\rangle \Big] \;,
\label{d2qq}
\eea
where $I^{1}_m$ is given by \eqrf{ds-q},
except for the constant terms $A$, $AA$, $AAA$ and $J$,
which do not contribute to the DSE.
The explicit expression for
$\langle I^{1}_mQ_n\rangle$ reads:
\bea
\langle I^{1}_mQ_n\rangle &=& \Gamma
_{mn'}^{(0)}(Q^2)\langle Q_{n'}Q_{n}\rangle +\Gamma
_{m'n'm}^{(0)}(AQ^2)A_{m'}\langle Q_{n'}Q_{n}\rangle
\nonumber\\
&&+\frac 12 \Gamma _{m'n'p'm}^{(0)}(A^2Q^2)A_{m'}A_{n'}
\langle Q_{p'}Q_{n}\rangle
\nonumber\\
&&+\frac 12 \Gamma _{mn'p'}^{(0)}(Q^3)\langle Q_{n'}Q_{p'}Q_{n}\rangle
+\frac 16 \Gamma _{mn'p'q'}^{(0)}(Q^4)
\langle Q_{n'}Q_{p'}Q_{q'}Q_{n}\rangle
\nonumber\\
&&+\frac 12 \Gamma _{m'n'p'm}^{(0)}(AQ^3)A_{m'}
\langle Q_{n'}Q_{p'}Q_{n}\rangle
\nonumber\\
&&+\Gamma _{m'n'm}^{(0)}(\ovl{C}CQ)
\langle \ovl{C}_{m'}C_{n'}Q_{n}\rangle
+\Gamma _{m'n'p'm}^{(0)}(\ovl{C}CAQ)A_{p'}
\langle \ovl{C}_{m'}C_{n'}Q_{n}\rangle
\nonumber\\
&&+\Gamma _{m'n'p'm}^{(0)}(\ovl{C}CQ^2) \langle
\ovl{C}_{m'}C_{n'}Q_{p'}Q_{n}\rangle \;.
\label{i1q}
\eea
We see that all terms in \eqrf{d2qq}
are $n$-point correlation functions,
or GFs, and can be expressed in terms of CGFs.
The relations between the GF and the CGF
are given by Eqs.\ (\ref{q-col}) and (\ref{qc-col})
in Appendix B. A higher-rank CGF can be related to
lower-rank CGFs and 1-PI vertices by various identities.
These identities are summarized in
Eqs.\ (\ref{cgq1}) and (\ref{cgq2}) in
Appendix B. For our further discussion
we also introduce the following
notions and conventions for the GF and CGF.
We denote the 2-, 3- and 4-point CGF for $Q$ 
$G_{mn}(Q^2)$, $G_{mnp}(Q^3)$ and $G_{mnpq}(Q^4)$, respectively.
For convenience, sometimes, we also use a short-hand notation
$(mn)\equiv G_{mn}(Q^2)$, $(mnp)\equiv G_{mnp}(Q^3)$, etc.
The 2-, 3- and 4-point CGFs for $Q$ and $\ovl{C}$/$C$ 
are denoted by 
$G_{mn}(\ovl{C}C)$, $G_{mnp}(\ovl{C}CQ)$ and
$G_{mnpq}(\ovl{C}CQ^2)$, respectively.
We also use the following
simplified notations: $([mn])\equiv G_{mn}(\ovl{C}C)$,
$([mn]p)\equiv G_{mnp}(\ovl{C}CQ)$, $([mn]pq)\equiv
G_{mnpq}(\ovl{C}CQ^2)$ etc.

We treat $A$ and $Q$ as two independent variables. In the last
step of the calculations, we always set $\langle Q\rangle$ and all
sources to zero. As it is only in the last step that 
$\langle Q\rangle$ is set to zero, the external sources and background
field $A$ are independent of each other in the intermediate
steps. This is the difference between our approach and that used
in Ref.\ \cite{soh86}. Thus, in deriving the DSE for the 2-point GF,
we can drop all terms in \eqrf{d2qq} that are proportional to
$\langle Q\rangle$ as we need not take further derivative
with respect to $\langle Q\rangle$. Note that
the correlation function $\langle Q_mQ_n\rangle$
becomes simply $G_{mn}(Q^2)$ after dropping
the $\langle Q\rangle$ term.

Following the identities
\bea
&&\frac{\delta ^2W}{\delta
J_p\delta J_q} \cdot \frac{\delta ^2\Gamma} {\delta \langle
Q_p\rangle \delta \langle Q_m\rangle }
=-\delta _{qm}\;, \nonumber\\
&&\frac{\delta ^2W}{i\delta J_p\delta J_q}=G_{pq}\;,
\eea
we obtain the DSE for 2-point GF from \eqrf{d2qq}:
\bee
\label{sch-dy0}
iG^{-1}_{mp}(Q^2)=iG^{-1}_{(0)mp}[Q^2]
+\Pi _{mp}(Q^2)\;,
\eee
where
\bea iG^{-1}_{mp}(Q^2)&=&\frac{\delta
^2\Gamma} {\delta \langle Q_p\rangle \delta \langle Q_m\rangle }
\;,\nonumber\\
iG^{-1}_{(0)mp}[Q^2]&=&\Gamma _{mp}^{(0)}(Q^2)
+\Gamma _{m'mp}^{(0)}(AQ^2)A_{m'}
+\frac 12 \Gamma _{m'n'mp}^{(0)}(A^2Q^2)A_{m'}A_{n'}
\;,\nonumber\\
\Pi _{mp}(Q^2)&=&\Pi ^{(1)}_{mp}+\Pi ^{(2)}_{mp}
\;,\nonumber\\
\Pi ^{(1)}_{mp}(Q^2)G_{pq}&=&
\frac 12 \Gamma _{mn'p'}^{(0)}(Q^3)\langle Q_{n'}Q_{p'}Q_{q}\rangle
+\frac 16 \Gamma _{mn'p'q'}^{(0)}(Q^4)
\langle Q_{n'}Q_{p'}Q_{q'}Q_{q}\rangle
\nonumber\\
&&+\frac 12 \Gamma _{m'n'p'm}^{(0)}(AQ^3)A_{m'}
\langle Q_{n'}Q_{p'}Q_{q}\rangle
\;,\nonumber\\
\Pi ^{(2)}_{mp}(Q^2)G_{pq}&=&
\Gamma _{m'n'm}^{(0)}(\ovl{C}CQ)
\langle \ovl{C}_{m'}C_{n'}Q_{q}\rangle
+\Gamma _{m'n'p'm}^{(0)}(\ovl{C}CAQ)A_{p'}
\langle \ovl{C}_{m'}C_{n'}Q_{q}\rangle
\nonumber\\
&&+\Gamma _{m'n'p'm}^{(0)}(\ovl{C}CQ^2) \langle
\ovl{C}_{m'}C_{n'}Q_{p'}Q_{q}\rangle \;.
\label{sch-dy-epl}
\eea
Note that $i\Pi _{pm}(Q^2)$ is the normal SE for $Q$;
$i\Pi ^{(1)}$ is the SE from the gluon loop and
$i\Pi ^{(2)}$ that from the gluon and ghost loops. Various bare
vertices $\Gamma ^{(0)}$ can be derived from
the classical action (\ref{s})
by taking the functional derivatives with respect to the
field expectation value. The results are summarized
in Appendix A.

The DSE (\ref{sch-dy0}) can be
written in an alternative form:
\bea
\label{sch-dy1}
iG^{-1}_{(0)mp}G_{pq}+\Pi _{mp}G_{pq}=i\delta _{mq}\;,\\
\label{sch-dy2}
G_{nm}iG^{-1}_{(0)mp}+G_{nm}\Pi _{mp}=i\delta _{np}\;.
\eea
where we omit the arguments $Q^2$ in $G_{mp}(Q^2)$,
$G^{-1}_{mp}(Q^2)$ and $i\Pi _{pm}(Q^2)$.
We take \eqrf{sch-dy1} as an example to expand and
analyse $\Pi _{mp}G_{pq}$. We know that
$\Pi _{mp}G_{pq}=\Pi ^{(1)}_{mp}G_{pq} +\Pi ^{(2)}_{mp}G_{pq}$
where $\Pi ^{(1,2)}_{mp}G_{pq}$ are given by \eqrf{sch-dy-epl}. Using
Eqs. (\ref{q-col}) and (\ref{cgq1}) from Appendix B,
we can expand $\Pi ^{(1)}_{mp}G_{pq}$ as follows:
\bea \Pi ^{(1)}_{mp}G_{pq}&=& \frac
12 \Gamma _{mn'p'}^{(0)}(Q^3)(n'p'q) +\frac 12 \Gamma
_{m'n'p'm}^{(0)}(AQ^3)A_{m'}(n'p'q)
\nonumber\\
&&+\frac 16 \Gamma _{mn'p'q'}^{(0)}(Q^4)
\Big[ i\Gamma _{r's't'u'}(Q^4)(r'n')(s'p')(t'q')(u'q)
\nonumber\\
&&+3i\Gamma _{r's't'}(Q^3)(r'n'q)(s'p')(t'q')
+3(n'p')(q'q)\Big]\;.
\label{pi1}
\eea
Note that the fraction in
front of each term is a symmetry factor, which is automatically
reproduced. In Fig.~1 we show the Feynman diagrams corresponding
to the above equation. The solid line there,  
with two dots at its ends, represents a
2-point full GF. The 3-point full GF is drawn as a shaded circle
with three solid-line legs, which have dots at their outer ends.
The 1-PI vertex is a shaded circle with all legs amputated; it has
dots on the circle that stands for points to which amputated legs
are attached. In the same way, we expand $\Pi ^{(2)}_{mp}G_{pq}$, 
which is associated with the ghost loop as
\bea
\Pi ^{(2)}_{mp}G_{pq}&=& \Gamma _{m'n'm}^{(0)}(\ovl{C}CQ)([m'n']q)
+\Gamma _{m'n'p'm}^{(0)}(\ovl{C}CAQ)A_{p'}([m'n']q)
\nonumber\\
&&+\Gamma _{m'n'p'm}^{(0)}(\ovl{C}CQ^2)
\Big\{([m'n']p'q)+([m'n'])(p'q)\Big\}
\nonumber\\
&=&\Gamma _{m'n'm}^{(0)}(\ovl{C}CQ)([m'n']q)
+\Gamma _{m'n'p'm}^{(0)}(\ovl{C}CAQ)A_{p'}([m'n']q)
\nonumber\\
&&+\Gamma _{m'n'p'm}^{(0)}(\ovl{C}CQ^2)\Big\{
i\Gamma _{r's't'}(\ovl{C}CQ^2)([m's']q)([r'n'])(t'p')
\nonumber\\
&&+i\Gamma _{r's't'}(\ovl{C}CQ^2)(m's')([r'n']q)(t'p')
\nonumber\\
&&+i\Gamma _{r's't'}(\ovl{C}CQ^2)([m's'])([r'n'])(t'p'q)
\nonumber\\
&&+i\Gamma _{r's't'u'}(\ovl{C}CQ^2)([m's'])([r'n'])(t'p')(u'q)
\nonumber\\
&&+([m'n'])(p'q)\Big\} \;.
\label{pi2}
\eea
In the expansion of Eqs.\ (\ref{pi1}) and (\ref{pi2}),
we have set $\langle Q\rangle$,
$\langle C\rangle$ and $\langle\ovl{C}\rangle$ to zero.
Feynman diagrams corresponding to \eqrf{pi2} are shown in Fig. 2.

Until now we have derived the DSE for the 2-point GF. In the
following we extend our analysis to the DSE for the 3-point GF.
Taking the derivative with respect to $\langle Q_l\rangle $ in
\eqrf{d2qq}, we obtain
\bea &&\frac{\delta}{\delta \langle
Q_l\rangle } \left[\frac{\delta ^2\Gamma} {\delta \langle
Q_p\rangle \delta \langle Q_m\rangle }\right]
=-i\Gamma _{lpn}(Q^3)\langle I^{1}_mQ_n\rangle\nonumber \\
&&-\Gamma _{pn}(Q^2)\Gamma _{ll'}(Q^2)
\Big( \langle I^{1}_mQ_nQ_{l'}\rangle -\langle Q_{l'}\rangle
\langle I^{1}_mQ_n\rangle \Big)\nonumber\\
&&+i\Gamma _{lpn}(Q^3)\langle Q_{n}\rangle\langle I^{1}_m\rangle
+\Gamma _{pn}(Q^2)\Gamma _{ll'}(Q^2)\nonumber \\
&&\cdot \Big[\langle I^{1}_m\rangle
\Big( \langle Q_{n}Q_{l'}\rangle
-\langle Q_{n}\rangle\langle Q_{l'}\rangle \Big)
+\langle Q_{n}\rangle \Big( \langle I^{1}_mQ_{l'}\rangle - \langle
I^{1}_m\rangle \langle Q_{l'}\rangle \Big) \Big] \;.
\eea
After setting $\langle Q\rangle =0$, we have
\bea \Gamma _{lpm}(Q^3)
&=&-i\Gamma _{lpn}(Q^3)\langle I^{1}_mQ_n\rangle -\Gamma
_{pn}(Q^2)\Gamma _{ll'}(Q^2)
\langle I^{1}_mQ_nQ_{l'}\rangle \nonumber\\
&&+\Gamma _{pn}(Q^2)\Gamma _{ll'}(Q^2) \langle I^{1'}_m\rangle
\langle Q_{n}Q_{l'}\rangle \;,
\eea
where $I^{1'}_m$ is as $I^{1}_m$ without the
$\langle Q\rangle$ terms.
From \eqrf{d2qq}, by dropping the term
$\langle Q_n\rangle \langle I^{1}_m\rangle $, the DSE becomes
$\langle I^{1}_mQ_n\rangle =i\delta _{mn}$. We apply it to the
above equation and obtain
\bee
\label{sd3p}
\langle I^{1}_mQ_nQ_{p}\rangle
=\langle Q_nQ_{p}\rangle \langle
I^{1'}_m\rangle \;.
\eee
Using \eqrf{i1q}, we explicitly write
$\langle I^{1}_mQ_nQ_{p}\rangle$ as follows:
\bea \langle I^{1}_mQ_nQ_{p}\rangle &=& \Gamma
_{mn'}^{(0)}(Q^2)\langle Q_{n'}Q_{n}Q_{p}\rangle +\Gamma
_{m'n'm}^{(0)}(AQ^2)A_{m'}
\langle Q_{n'}Q_{n}Q_{p}\rangle \nonumber\\
&&+\frac 12 \Gamma _{m'n'p'm}^{(0)}(A^2Q^2)A_{m'}A_{n'}
\langle Q_{p'}Q_{n}Q_{p}\rangle \nonumber\\
&&+\frac 12 \Gamma _{mn'p'}^{(0)}(Q^3)
\langle Q_{n'}Q_{p'}Q_{n}Q_{p}\rangle
+\frac 16 \Gamma _{mn'p'q'}^{(0)}(Q^4)
\langle Q_{n'}Q_{p'}Q_{q'}Q_{n}Q_{p}\rangle \nonumber\\
&&+\frac 12 \Gamma _{m'n'p'm}^{(0)}(AQ^3)A_{m'}
\langle Q_{n'}Q_{p'}Q_{n}Q_{p}\rangle \nonumber\\
&&+\Gamma _{m'n'm}^{(0)}(\ovl{C}CQ)
\langle \ovl{C}_{m'}C_{n'}Q_{n}Q_{p}\rangle \nonumber\\
&&+\Gamma _{m'n'p'm}^{(0)}(\ovl{C}CAQ)A_{p'}
\langle \ovl{C}_{m'}C_{n'}Q_{n}Q_{p}\rangle \nonumber\\
&&+\Gamma _{m'n'p'm}^{(0)}(\ovl{C}CQ^2) \langle
\ovl{C}_{m'}C_{n'}Q_{p'}Q_{n}Q_{p}\rangle \;. \eea
We can prove that the disconnected part of
$\langle I^{1}_mQ_nQ_{p}\rangle$
cancels $\langle Q_nQ_{p}\rangle \langle
I^{1'}_m\rangle$, hence we write \eqrf{sd3p} as
\bee \label{sd3p1}
iG^{-1}_{(0)mn'}[Q^2]G_{n'np}(Q^3)=-\Gamma _{mnp} \;,
\eee
where the 1-PI vertex $\Gamma _{mnp}$ is defined by
\bea \Gamma _{mnp} &=&\frac 12 \Gamma _{mn'p'}^{(0)}(Q^3) \langle
Q_{n'}Q_{p'}Q_{n}Q_{p}\rangle _c +\frac 12 \Gamma
_{m'n'p'm}^{(0)}(AQ^3)A_{m'}
\langle Q_{n'}Q_{p'}Q_{n}Q_{p}\rangle _c\nonumber\\
&&+\frac 16 \Gamma _{mn'p'q'}^{(0)}(Q^4)
\langle Q_{n'}Q_{p'}Q_{q'}Q_{n}Q_{p}\rangle _c\nonumber\\
&&+\Gamma _{m'n'm}^{(0)}(\ovl{C}CQ)
\langle \ovl{C}_{m'}C_{n'}Q_{n}Q_{p}\rangle _c\nonumber\\
&&+\Gamma _{m'n'p'm}^{(0)}(\ovl{C}CAQ)A_{p'}
\langle \ovl{C}_{m'}C_{n'}Q_{n}Q_{p}\rangle _c\nonumber\\
&&+\Gamma _{m'n'p'm}^{(0)}(\ovl{C}CQ^2) \langle
\ovl{C}_{m'}C_{n'}Q_{p'}Q_{n}Q_{p}\rangle _c \;, \label{gamma-c}
\eea
and the subscript $c$ stands for the connected part.

We can expand $\Gamma _{mnp}$ term by term.
The first term $\frac 12 \Gamma _{mn'p'}^{(0)}(Q^3) \langle
Q_{n'}Q_{p'}Q_{n}Q_{p}\rangle _c$ in \eqrf{gamma-c} can be
obtained by expressing $\langle Q_{n'}Q_{p'}Q_{n}Q_{p}\rangle $ by
the CGFs of the lower rank. This relation is given by
\eqrf{gf4q} in Appendix B. We can identify in this relation
disconnected GFs $(n'p')(np)$. After dropping it, we get the
connected part $\langle Q_{n'}Q_{p'}Q_{n}Q_{p}\rangle _c$. The
corresponding Feynman diagrams are shown in Fig.~3. The second
term in \eqrf{gamma-c} is the same as the first one, except that
there is an additional $A$ field attached to the bare vertex. The
third, fourth and fifth terms can be expressed in terms of the
lower rank CGFs. The resulting relations are given respectively
by Eqs.\ (\ref{gf5q1}-\ref{gf5c})
in Appendix B.
Their corresponding Feynman diagrams are shown in Figs.~4-6.

In this section, we have derived the DSE for the 2- and 3-point
GFs. We strictly stick to the functional definition of the
1-PI vertex and the CGF and use the DeWitt notation.
Finally, the relations between CGFs and 1-PI vertices were
recursively applied to express a higher rank CGF in terms of the
lower rank ones and 1-PI vertices. The current approach has the
advantage that the non-local terms can be treated in the
same way as the local ones. 
The difference between the local vertex and the non-local one
is that the former has a sufficient number of $\delta $-functions
to ensure that vertex is at the same space-time point.
Another advantage of the current approach is that it can
produce all needed Feynman diagrams automatically.
Hence it is easy to implement our approach in a
computer algorithm, which can automatically generate Feynman diagrams
for a given process.

\section{Dyson-Schwinger Equation in Closed-Time-Path Formalism}

The non-equilibrium dynamics is usually described in the
CTP formalism \cite{schwinger,keldysh}.
In this section, we will formulate the DSE in
this formalism. 
The generating functional
$Z[J,\xi,\ovl{\xi}]$ in the CTP formalism reads:
\bea
Z[A_{\pm},J_{\pm},\xi _{\pm},\ovl{\xi}_{\pm}] &=&\int
[dQ_+][dQ_-][dC_+][dC_-][d\ovl{C}_+][d\ovl{C}_-]
\nonumber\\
&&\cdot \exp \{iS_+-iS_-+iK(A_{\pm},Q_{\pm})\}
\nonumber\\
S_+&\equiv &S(A_+,Q_+,C_+,\ovl{C}_+,J_+,\xi _+,\ovl{\xi}_+)
\nonumber\\
S_-&\equiv &S(A_-,Q_-,C_-,\ovl{C}_-,J_-,\xi _-,\ovl{\xi}_-) \;,
\eea
where the classical action $S$ of BG-QCD is given by \eqrf{s}.
We denote the total action as $S_{CTP}=S_+ - S_-$. We have omitted
the kernel $K(A_{\pm},Q_{\pm})$. This is because, as we
will see in the next section, the kernel $K$ can be put into the
boundary condition. We can treat the $+$ and $-$
quantities as if they were independent.
Then, we can derive the Feynman rules according
to $S_{CTP}$. The main difference between the 
tree-level vertices obtained from $S_{CTP}$ and from  $S$ lies in
that there are negative-type vertices (all time arguments are on
the $-$ branch) besides the ordinary positive ones. The vertices of
the negative and positive types are the same, except that they have
opposite sign. At the tree level, there is also no vertex of
mixed type with both positive and negative time arguments.

The 2-point GF for $Q$ is defined by
$G=\langle T_{\rm P}Q(x_1)Q(x_2)\rangle $, 
where $T_{\rm P}$ indicates a path-ordered product along the CTP.
For simplicity we suppress the Lorentz and
colour indices of $Q$ and restore them when necessary.
There are four types of 2-point GFs characterized by the time-branch
($+$ or $-$). This can be explicitly written as a matrix:
\bea
G&=&\left( \ba{cc}
\langle {\rm T}Q(x_1)Q(x_2)\rangle&\langle Q(x_2)Q(x_1)\rangle\\
\langle Q(x_1)Q(x_2)\rangle&\langle {\rm T}^*Q(x_1)Q(x_2)\rangle \ea
\right)\non
&=& \left( \ba{cc}
G^{++}&G^{+-}\\
G^{-+}&G^{--}
\ea
\right)
=\left( \ba{cc}
G^{F}&G^{<}\\
G^{>}&G^{\ovl{F}}
\ea
\right) \nonumber \;,
\eea
where $T^*$ denotes the counter-time-ordered
product; $G^{++}$ ($G^{F}$) is the ordinary GF
and its time arguments $t_1$ and $t_2$
are on the positive time branch;
$G^{--}$ ($G^{\ovl{F}}$) is the counter-time-ordered 
or anticausal GF, with both time arguments on the negative time branch;
$G^{+-}$ ($G^<$) and $G^{-+}$ ($G^>$) are correlation functions with
time arguments on different branches.
The SE has a similar form:
\bee
\Pi = \left( \ba{cc}\Pi ^{F}&\Pi ^{<}\\
\Pi ^{>}&\Pi ^{\ovl{F}}\ea \right)\;.
\eee
The GF and the SE can be expressed in the so-called
physical representation by using the following unitary
transformation:
\bea
U\left( \ba{cc}\Pi ^{F}&\Pi ^{<}\\
\Pi ^{>}&\Pi ^{\ovl{F}}\ea \right) U^{-1}
=\left( \ba{cc}\Pi ^{C}&\Pi ^{R}\\
\Pi ^{A}&0\ea \right)\;, \non
U\left( \ba{cc}G^{F}&G^{<}\\
G^{>}&G^{\ovl{F}}\ea \right) U^{-1}
=\left( \ba{cc}0&G^{A}\\
G^{R}&G^{C}\ea \right)\;,
\label{physical}
\eea
where
\bee
U=\frac{1}{\sqrt{2}}\left(
\ba{cc}1&-1\\1&1\ea \right),\,\,\,\,
U^{-1}=\frac{1}{\sqrt{2}}
\left( \ba{cc}1&1\\-1&1\ea\right)\;.
\eee
Writing the transformation (\ref{physical})
explicitly, we obtain the following relations:
\bea
&&G^A=G^F-G^>=G^<-G^{\ovl{F}},\,\,\,\, \Pi ^A=\Pi ^F+\Pi
^>=-\Pi ^<-\Pi ^{\ovl{F}}\;,
\nonumber\\
&&G^R=G^F-G^<=G^>-G^{\ovl{F}},\,\,\,\,
\Pi ^R=\Pi ^F+\Pi ^<=-\Pi ^>-\Pi ^{\ovl{F}}\;,
\nonumber\\
&&G^C=G^F+G^{\ovl{F}}=G^<+G^>,\,\,\,\, \Pi ^C=\Pi ^F+\Pi
^{\ovl{F}}=-\Pi ^<-\Pi ^> \;.
\eea
with $A$, $R$ and $C$ denoting the advanced, retarded and
homogeneous GF or SF. Note that there is an additional negative
sign for $\Pi ^{>,<}$ with respect to $G^{>,<}$.
The reason is as follows:
the SE tensor $\Pi$ is a 1-PI 2-point
GF with external legs amputated, and the
two time arguments of $\Pi ^{>,<}$
are on different CTP branches. We also know that 
negative and positive type vertices differ in sign.
Therefore, $\Pi ^{>,<}$ has an additional negative sign
relative to $G^{>,<}$. Such a case, however, does not occur for
$\Pi ^{\ovl{F}}$, because both of its time arguments are on
the negative branch; they thus contribute
with the same negative signs, which cancel.

Keeping these Feynman rules in mind, we can write the
DSE (\ref{sch-dy1}) and (\ref{sch-dy2})
in the ordinary representation as
\bea
&&\left( \ba{cc}{\mathcal D}+\Pi ^+&0\\
0&-{\mathcal D}-\Pi ^+ \ea \right)
_{\mu\nu ;ab}(x_1)
\left( \ba{cc}G^{F}&G^{<}\\G^{>}&G^{\ovl{F}}\ea \right)
_{\nu\lambda ;bc}(x_1,x_2)\nonumber \\
&&=ig_{\mu\lambda}\delta _{ac}\delta(x_1-x_2)
\left( \ba{cc}1&0\\0&1\ea \right)
\nonumber \\
&&-\int d^4x'
\left( \ba{cc}\Pi ^{F}&\Pi ^{<}\\\Pi ^{>}&\Pi ^{\ovl{F}}\ea \right)
_{\mu\nu ;ab}(x_1,x')
\left( \ba{cc}G^{F}&G^{<}\\G^{>}&G^{\ovl{F}}\ea \right)
_{\nu\lambda ;bc}(x',x_2) \;,
\label{sch-dy3}
\eea
\bea
&&\left( \ba{cc}G^{F}&G^{<}\\G^{>}&G^{\ovl{F}}\ea \right)
_{\mu\nu ; ab}(x_1,x_2)
\left( \ba{cc}{\mathcal D}^{\dagger}+\Pi ^+&
0\\0&-{\mathcal D}^{\dagger}-\Pi ^+
\ea \right)_{\nu\lambda ; bc}(x_2)
\nonumber \\
&&=ig_{\mu\lambda}\delta _{ac}\delta(x_1-x_2)
\left( \ba{cc}1&0\\0&1\ea \right)
\nonumber \\
&&-\int d^4x'
\left( \ba{cc}G^{F}&G^{<}\\G^{>}&G^{\ovl{F}}\ea \right)
_{\mu\nu ; ab}(x_1,x')
\left( \ba{cc}\Pi ^{F}&\Pi ^{<}\\\Pi ^{>}&\Pi ^{\ovl{F}}\ea \right)
_{\nu\lambda ; bc}(x',x_2)\;,
\label{sch-dy4}
\eea
or, in the physical representation, as
\bea
&&\left( \ba{cc}{\mathcal D}+\Pi ^+&0\\
0&-{\mathcal D}-\Pi ^+ \ea \right)
_{\mu\nu ;ab}(x_1)
\left( \ba{cc}G^{R}&G^{C}\\
0&G^{A}\ea \right)_{\nu\lambda;bc}(x_1,x_2)\nonumber \\
&&=ig_{\mu\lambda}\delta _{ac}\delta(x_1-x_2)
\left( \ba{cc}1&0\\0&1\ea \right)\nonumber \\
&&-\int d^4x'
\left( \ba{cc}\Pi ^{C}&\Pi ^{R}\\\Pi ^{A}&0\ea \right)
_{\mu\nu ;ab}(x_1,x')
\left( \ba{cc}0&G^{A}\\G^{R}&G^{C}\ea \right)
_{\nu\lambda ;bc}(x',x_2) \;,
\label{sch-dy5}
\eea
\bea
&&\left( \ba{cc}G^{A}&0\\G^{C}&G^{R}\ea \right)_{\mu\nu;ab}(x_1,x_2)
\left( \ba{cc}{\mathcal D}^{\dagger}+\Pi ^+&
0\\0&-{\mathcal D}^{\dagger}-\Pi ^+
\ea \right)_{\nu\lambda ; bc}(x_2)
\nonumber \\
&&=ig_{\mu\lambda}\delta _{ac}\delta(x_1-x_2)
\left( \ba{cc}1&0\\0&1\ea \right)\nonumber \\
&&-\int d^4x' \left( \ba{cc}0&G^{A}\\G^{R}&G^{C}\ea \right)
_{\mu\nu;ab}(x_1,x') \left( \ba{cc}\Pi ^{C}&\Pi ^{R}\\\Pi
^{A}&0\ea \right) _{\nu\lambda;bc}(x',x_2)\;.
\label{sch-dy6}
\eea
In \eqrfm{sch-dy3}{sch-dy6}, $(x_1)$ and $(x_1,x_2)$ stand for
space-time arguments for the functions in front of them.
The Lorentz and colour indices are written as subscripts.
As an alternative way of presenting the above equations,
the local term $\pm \Pi ^+$ can be absorbed into the
definition of $\Pi ^{F/\ovl{F}}$.
In the Feynman gauge ($\alpha =1$) the differential
operators ${\mathcal D}$ and ${\mathcal D}^{\dagger}$
are defined as:
\bea
{\mathcal D}^{hi}_{\rho\sigma}
&=&g_{\rho\sigma}D_{\mu}^{ha}[A]D_{\mu}^{ai}[A]
-D_{\sigma}^{ha}[A]D_{\rho}^{ai}[A]\nonumber\\
&&+\frac 1{\alpha}D_{\rho}^{ha}[A]D_{\sigma}^{ai}[A]
+gf^{hai}F^a_{\rho\sigma}[A]\nonumber\\
&=&g_{\rho\sigma}D_{\mu}^{ha}[A]D_{\mu}^{ai}[A]
+2gf^{hai}F^a_{\rho\sigma}[A] \;,
\label{d1}
\eea
and
\bea
{\mathcal D}^{\dagger;hi}_{\rho\sigma}
&=&g_{\rho\sigma}D_{\mu}^{\dagger;ha}[A]D_{\mu}^{\dagger;ai}[A]
+2gf^{hai}F^a_{\rho\sigma}[A] \;,
\label{d2}
\eea
where
$D_{\mu}^{ha}[A] =\partial _{\mu}\delta ^{ha}
+gf^{hba}A_{\mu}^b$
is the covariant derivative in the
adjoint representation and
$D_{\mu}^{\dagger;ha}=
\stackrel{\leftarrow}{\partial }_{\mu}
\delta ^{ha}-gf^{hba}A_{\mu}^b$
is the conjugate covariant derivative,
where the differential operator acts to the left.
We note that \eqrfm{sch-dy3}{sch-dy6}
are independent of the gauge parameter
$\a $, owing to the gauge conditions:
$D^{ij}_{\mu}[A(x_1)]G_{\mu\nu}^{jk}(x_1,x_2)=0$ and
$G^{ij}_{\mu\nu}(x_1,x_2)D_{\nu}^{\dagger ;jk}[A(x_2)]=0$.

Some comments about Eqs.\ (\ref{sch-dy3}-\ref{d2}) 
are in order. First, we recall that \eqrfm{d1}{d2}
come from the second line of \eqrf{sch-dy-epl}.
We write it down explicitly:
\bea
iG^{-1;hi}_{(0);\rho\sigma}(x,y)&=&\Gamma
_{mp}^{(0)}(Q^2)+\Gamma _{m'mp}^{(0)}(AQ^2)A_{m'}
+\frac 12 \Gamma _{m'n'mp}^{(0)}(A^2Q^2)A_{m'}A_{n'}
\nonumber\\
&=&\Big\{ g_{\rho\sigma}D_{\mu}^{ha}[A(x)]D_{\mu}^{ai}[A(x)]
+2gf^{hai}F^a_{\rho\sigma}[A(x)]\Big\}\delta ^4(x-y) \;,
\label{d0}
\eea
where labels $m$ and $p$ stand for a group of indices:
$m=(x,\rho,h)$ and $p=(y,\sigma,i)$. Multiplying
\eqrf{d0} by $G_{\sigma\xi}^{ij}(y,z)$ and
integrating over $y$, \eqrf{d0} gives
\bea
&&\int d^4y iG^{-1;hi}_{(0);\rho\sigma}(x,y)
G_{\sigma\xi}^{ij}(y,z)\nonumber\\
\nonumber\\
&&=\Big\{ g_{\rho\sigma}D_{\mu}^{ha}[A(x)]D_{\mu}^{ai}[A(x)]
+2gf^{hai}F^a_{\rho\sigma}[A(x)]\Big\}
\int d^4y \delta ^4(x-y)G_{\sigma\xi}^{ij}(y,z)
\nonumber\\
&&=\Big\{ g_{\rho\sigma}D_{\mu}^{ha}[A(x)]D_{\mu}^{ai}[A(x)]
+2gf^{hai}F^a_{\rho\sigma}[A(x)]\Big\} G_{\sigma\xi}^{ij}(x,z)
\nonumber\\
&&={\mathcal D}^{hi}_{\rho\sigma}(x) G_{\sigma\xi}^{ij}(x,z) \;.
\eea
Multiplying \eqrf{d0} by $G_{\xi\rho}^{jh}(z,x)$ and
integrating over $x$, it becomes:
\bea
&&\int d^4x
G_{\xi\rho}^{jh}(z,x)
iG^{-1;hi}_{(0);\rho\sigma}(x,y)\nonumber\\
&&=\int d^4x G_{\xi\rho}^{jh}(z,x)
\Big\{ g_{\rho\sigma}D_{\mu}^{ha}[A(x)]D_{\mu}^{ai}[A(x)]
+2gf^{hai}F^a_{\rho\sigma}[A(x)]\Big\}\delta ^4(x-y)\nonumber\\
&&=\int d^4x G_{\xi\rho}^{jh}(z,x)
\Big\{ g_{\rho\sigma}D_{\mu}^{\dagger;ha}[A(x)]
D_{\mu}^{\dagger;ai}[A(x)]
+2gf^{hai}F^a_{\rho\sigma}[A(x)]\Big\}
\delta ^4(x-y)\nonumber\\
&&=G_{\xi\rho}^{jh}(z,y)
\Big\{ g_{\rho\sigma}D_{\mu}^{\dagger;ha}[A(y)]
D_{\mu}^{\dagger;ai}[A(y)]
+2gf^{hai}F^a_{\rho\sigma}[A(y)]\Big\}\nonumber\\
&&=G_{\xi\rho}^{jh}(z,y)
{\mathcal D}^{\dagger;hi}_{\rho\sigma}(y)
\;, \eea
where the differential operator $\partial$ acting on the
$\delta $-function in the second line is changed to
$-\stackrel{\leftarrow}{\partial}$
in the third line. We notice that
$G^{-1;hi}_{(0);\rho\sigma}(x,y)$ in \eqrf{d0}
should be understood as $G^{-1;hi}_{(0);\rho\sigma}(x_+,y_+)$,
where $x_+$ and $y_+$ are on the positive time branch.
In addition, since $G^{-1;hi}_{(0);\rho\sigma}(x,y)$ is associated with
vertices $\Gamma _{mp}^{(0)}(Q^2)$, $\Gamma _{m'mp}^{(0)}(AQ^2)$
and $\Gamma _{m'n'mp}^{(0)}(A^2Q^2)$,
$G^{-1;hi}_{(0);\rho\sigma}(x_-,y_-)$ should have an additional
negative sign. This is the reason why there is a negative sign
before ${\mathcal D}$ and ${\mathcal D}^{\dagger}$ in
\eqrfm{sch-dy3}{sch-dy6}.

Of particular relevance to our further discussion are the
specific matrix elements originating from Eqs.\ (\ref{sch-dy5})
and (\ref{sch-dy6}). The equations corresponding to the
upper-right element of (\ref{sch-dy5}) and  to the lower-left
element of (\ref{sch-dy6}) are:
\bea
&&\Big[ {\mathcal D}(x_1)+\Pi ^+(x_1)\Big]G^C(x_1,x_2)\nonumber\\
&&=-\int d^4x'\Big[ \Pi ^C(x_1,x')G^A(x',x_2)
+\Pi ^{R}(x_1,x')G^{C}(x',x_2)\Big] \;,
\label{sch-dy-c1}\\
&&G^C(x_1,x_2)\Big[ {\mathcal D}^{\dagger}(x_2)+\Pi ^+(x_2)\Big]
\nonumber\\
&&=-\int d^4x'\Big[ G^R(x_1,x')\Pi ^C(x',x_2) +G^{C}(x_1,x')\Pi
^{A}(x',x_2)\Big] \;,
\label{sch-dy-c2}
\eea
where, for simplicity,
we suppress colour and Lorentz indices in
$G$ and $\Pi$, and in particular we regard them
as matrices in colour space.
We note that the local SE tensor $\Pi ^+$ can be
absorbed into $\Pi ^{R}$ and $\Pi ^{A}$ to make
Eqs.\ (\ref{sch-dy-c1}) and (\ref{sch-dy-c2}) more compact.

\section{Discussions of the non-local source}

So far we have not discussed the role of the non-local source
term $K(A_{\pm},Q_{\pm})$ in the DSE. The reason is that
$K(A_{\pm},Q_{\pm})$ is non-zero only at the initial time and
hence can be put to the boundary condition.
To illustrate the role of $K$, we consider a simple example of a
non-interacting massive scalar field.
The classical Lagrangian density is
${\mathcal L}=\frac 12\partial _{\mu} \phi\partial _{\mu}\phi
-\frac 12m^2\phi ^2$. In the CTP approach
the role of the initial density matrix is taken by the
non-local source kernel $K(\phi _{\pm})$.
$K(\phi _{\pm})$ can be expanded
into a functional Taylor series in $\phi$.
The linear term $K_a\phi _a$ of this series can be
absorbed into the source term $J_a\phi _a$.
The lowest order term,
which does have an effect on the dynamics,
is the square term $K_{ab}\phi _a\phi _b$.
In this heuristic argument, we neglect
higher order terms in this expansion.
Using this approximation the
DSE has a very simple form:
\bee
\Big[ \delta ^4(x-y)(-\partial _y^2-m^2)
+K(x,y)\Big] \eta \Delta (y,z)=\delta ^4(x-z)\;,
\eee
where $\eta ={\rm diag} (1,-1)$; $i\Delta (y,z)$ is a 2-point
CTP-form GF which is a $2\times 2$ matrix and $K(x,y)$ is a
CTP-form matrix. The solution of the above equation can be written as
\bea
\Delta (x,y)&=&\Delta ^{(0)}(x,y) -\Delta
^{(0)}(x,u_1)\eta K(u_1,u_2)\eta\Delta ^{(0)}(u_2,y)
\nonumber\\
&&+\Delta ^{(0)}(x,u_1)\eta K(u_1,u_2)\eta
\Delta ^{(0)}(u_2,u_3)\eta K(u_3,u_4)\eta\Delta ^{(0)}(u_4,y)
\nonumber\\
&&-\cdots \;,
\label{deltsol}
\eea
where $\Delta ^{(0)}(x,y)$ is a solution of
the inhomogeneous equation
\bee
(-\partial _y^2-m^2) \eta \Delta ^{(0)}(x,y)
=\delta ^4(x-y) \;,
\eee
which we call a full propagator. The propagator
$\Delta ^{(0)}(x,y)$ can be written as the sum of the homogeneous
solution and an inhomogeneous one:
$\Delta ^{(0)}=\Delta ^{(0)}_{in}+\Delta ^{(0)}_{hom}$,
where $i\Delta ^{(0)}_{in}(x-y)$
is the Feynman propagator and $\Delta ^{(0)}_{hom}(x-y)$ is the
solution of the homogeneous equation
\bee
(-\partial _y^2-m^2)
\eta \Delta _{hom}(x,y)=0\;.
\eee
In momentum space, these propagators become
\bea
\Delta ^{(0)}(p,q)&=&\Delta ^{(0)}(p)\delta ^4(p-q)\;,
\nonumber\\
\Delta ^{(0)}(p)&=&\Delta ^{(0)}_{in}(p)
+\Delta ^{(0)}_{hom}(p) \;,
\eea
where $\Delta ^{(0)}_{in}(p)$
and $\Delta ^{(0)}_{hom}(p)$ are given by
\bea
\Delta ^{(0)}_{in}(p)&=&
\left ( \ba{cc}\frac{1}{p^2-m^2+i\epsilon }&0\\0&
\frac{-1}{p^2-m^2-i\epsilon }\ea \right )\;,\nonumber \\
\Delta ^{(0)}_{hom}(p)&=&-2\pi i\delta (p^2-m^2)
\left ( \ba{cc}g^{(0)}({\mathbf p})&\theta (-p^0)
+g^{(0)}({\mathbf p})\\ \theta (p^0)
+g^{(0)}({\mathbf p})&g^{(0)}({\mathbf p})
\ea \right )\nonumber \\
&=&-2\pi i\delta (p^2-m^2)\Big[ \lambda _1
+g^{(0)}({\mathbf p})\lambda _2\Big]\;,
\eea
with
\bee
\lambda _1=\left( \ba{cc}0&\theta (-p^0)
\\ \theta (p^0)&0 \ea \right)\;,\;\;\;\;
\lambda _2=\left( \ba{cc}1&1\\1&1 \ea \right)\;.
\eee

In our further discussion
we assume that the kernel $K(x_1,x_2)$
is given in the following form: 
\bee
K^{ij}(x_1,x_2) =\frac 1{(2\pi)^3} K^{ij}({\mathbf
x}_1-{\mathbf x}_2) \delta (x_1^0-t_0^i)\delta (x_2^0-t_0^j)\;,
\eee
where $i,j=\pm $ and $t_0^+$ and $t_0^-$ are,
respectively, the starting and ending points of the CTP.
The kernel $K^{ij}(x_1,x_2)$ is translationally
invariant in space, and its Fourier transform reads
\bea
K^{ij}(k_1,k_2)&=&\int
d^4x_1d^4x_2 e^{ik_1x_1}e^{-ik_2x_2}
K^{ij}(x_1,x_2)\nonumber\\
&=&K^{ij}({\mathbf k}_1)
\delta ^3({\mathbf k}_1-{\mathbf k}_2)\;,\nonumber\\
K^{ij}({\mathbf k}_1)&\equiv &\int d^3y
K^{ij}({\mathbf y})e^{-i{\mathbf k}_1\cdot {\mathbf y}}\;, 
\eea
where we use the same symbol $K$ to denote the kernel in the
coordinate and in the momentum space.

Let us now consider the full propagators
$\Delta ^{(0)}(x,u_1)$ and $\Delta ^{(0)}(u_n,x)$ appearing in
\eqrf{deltsol} at the {\it leading} and at the {\it end} of
each term. The values of $u_1^0$ and $u_n^0$ are $t_0^+$ or
$t_0^-$. If one assumes that
$t_0^+\ra -\infty ^+$ and $t_0^-=-\infty ^-$ then
$\Delta ^{(0);++}(x,u_1)=\Delta ^{(0);>}(x,u_1)$ since $x^0>t_0^+$
($x^0$ is definite). Similarly, we have:
$\Delta ^{(0);+-}(x,u_1)=\Delta ^{(0);<}(x,u_1)$,
$\Delta ^{(0);-+}(x,u_1)=\Delta ^{(0);>}(x,u_1)$
and $\Delta ^{(0);--}(x,u_1)=\Delta ^{(0);<}(x,u_1)$.
The $++$, $+-$, $-+$ and $--$ components of
$\Delta ^{(0)}(u_n,x)$ are of type $<$, $<$,
$>$ and $>$, respectively. In momentum space we have
\bea
\Delta ^{(0)}_h(p)&=&-2\pi i\delta (p^2-m^2) \Big[
\lambda _3 + g^{(0)}({\mathbf p})\lambda _2 \Big]\;,
\nonumber\\
\Delta ^{(0)}_e(q)&=&-2\pi i\delta (q^2-m^2)
\Big[ \lambda _4 + g^{(0)}({\mathbf q})\lambda _2 \Big]\;,
\eea
where the subscripts
$h$ and $e$ denote the {\it leading}
and the {\it end} propagator,
respectively, and $\lambda _3$ and $\lambda _4$ 
are defined as 
\bee
\lambda _3=\left( \ba{cc}\theta (p^0)&\theta (-p^0)
\\ \theta (p^0)&\theta (-p^0) \ea \right)\;,\;\;\;\;
\lambda _4=\left( \ba{cc}\theta (-q^0)&\theta (-q^0)
\\\theta (q^0)&\theta (q^0) \ea \right)\;.
\eee
The propagators $\Delta^0(u_i,u_{i+1})$ appearing in
\eqrf{deltsol} have both of their time
variables pinched at $t_0$.
Thus, they are not bound by the above arguments
and their four components are equal.
Finally, in momentum space, the full
propagator (\ref{deltsol}) becomes
\bee
\label{dpq}
\Delta (p,q)=\Delta ^{(0)}(p)\delta ^4(p-q) -\Delta
^{(0)}_{h}(p){\mathcal K}({\mathbf p}) 
\Delta ^{(0)}_{e}(q)\delta ^3({\mathbf p}-{\mathbf q}) \;, 
\eee 
where ${\mathcal K}({\mathbf
p})$ is given by 
\bea {\mathcal K}({\mathbf p})&=&\eta K({\mathbf
p})\eta -\eta K({\mathbf p})\eta \tilde{\Delta}^{(0)}({\mathbf p})
\lambda _2 \eta K({\mathbf p})\eta \nonumber\\
&&+\eta K({\mathbf p})\eta \tilde{\Delta}^{(0)}({\mathbf p}) 
\lambda _2\eta K({\mathbf p})\eta \tilde{\Delta}^{(0)}({\mathbf p}) 
\lambda _2\eta K({\mathbf p})\eta -\cdots \;,
\label{deltsol2}
\eea
and $\tilde{\Delta}^{(0)}({\mathbf p})= \int dp^0\Delta
^{(0)}(p)$. Collecting the sum in the above equation, we obtain:
\bea
{\mathcal K}^{++}({\mathbf p})&=&C \Big[
(1+AB_2)K^{++}({\mathbf p})+AB_1K^{+-}({\mathbf p})\Big]\;,
\nonumber\\
{\mathcal K}^{+-}({\mathbf p})&=&C
\Big[ -(1+AB_1)K^{+-}({\mathbf p})-AB_2K^{++}({\mathbf p})\Big]\;,
\nonumber\\
{\mathcal K}^{-+}({\mathbf p})&=&C
\Big[ -(1+AB_2)K^{-+}({\mathbf p})-AB_1K^{--}({\mathbf p})\Big]\;,
\nonumber\\
{\mathcal K}^{--}({\mathbf p})&=&C
\Big[ (1+AB_1)K^{--}({\mathbf p})
+AB_2K^{-+}({\mathbf p})\Big]\;,
\eea
where $A=\tilde{\Delta }^{(0)}({\mathbf p})$,
$B_1=K^{++}({\mathbf p})-K^{-+}({\mathbf p})$,
$B_2=K^{--}({\mathbf p})-K^{+-}({\mathbf p})$
and $C=1/[1+A(B_1+B_2)]$.

There are three types of contributions to the second term of
\eqrf{dpq}: $\lambda _3{\mathcal K}\lambda _4$,
$\lambda _3{\mathcal K}\lambda _2 +\lambda _2{\mathcal K}\lambda _4$, 
and $\lambda _2{\mathcal K}\lambda _2$.
We denote them as $I_1$, $I_2$ and $I_3$ respectively:
\bea
&&I_1=I_0 [{\mathcal K}^{++}\delta (p^0+q^0)+
{\mathcal K}^{+-}\delta (p^0-q^0)]\lambda _2\;,\nonumber\\
&&I_2=I_0 g^{(0)}({\mathbf p})({\mathcal K}^{++}+{\mathcal K}^{+-})
[\delta (p^0-q^0)+ \delta (p^0+q^0)]\lambda _2\;,\nonumber\\
&&I_3=I_0 [g^{(0)}({\mathbf p})]^2 ({\mathcal K}^{++}+{\mathcal
K}^{+-})[\delta (p^0-q^0)+ \delta (p^0+q^0)]\lambda _2 \;,
\label{delt}
\eea
where $I_0$ is defined by
\bee I_0=8\pi ^2\delta (p^2-m^2)\delta ^3({\mathbf
p}-{\mathbf q}) \frac 1{2E_{\mathbf p}}
\eee
with $E_{\mathbf p}=\sqrt{{\mathbf p}^2+m^2}$.
In \eqrf{delt} we have used: 
$K^{++}=K^{--}$ and $K^{+-}=K^{-+}$, 
which implies that ${\mathcal K}^{++}={\mathcal K}^{--}$ and
${\mathcal K}^{+-}={\mathcal K}^{-+}$. 
We have also used the following formula
\bee
\delta (p^2-m^2)\delta (q^2-m^2) 
\delta ^3(\mathbf{p}-\mathbf{q}) 
=\delta (p^2-m^2)\frac{1}{2E_{\mathbf{p}}}
[\delta (p^0-q^0)+\delta (p^0+q^0)] \;. 
\eee
With \eqrf{delt}, \eqrf{dpq} becomes
\bea
\Delta (p,q)&=&\Delta ^{(0)}(p)\delta ^4(p-q)
+I_1+I_2+I_3\nonumber\\
&=&\bigg\{\;\Delta ^{(0)}_{in}(p)-2\pi i\delta (p^2-m^2)
\Big[\; \lambda _1 + f({\mathbf p})
\lambda _2\;\Big]\;\bigg\}\;\delta ^4(p-q)\non
&&-2\pi i\delta (p^2-m^2)f'({\mathbf p}) \lambda _2 
\delta (p_0+q_0)\delta ^3({\mathbf p}-{\mathbf q})\;,
\label{delt1}
\eea
where $f({\mathbf p})$ and $f'({\mathbf p})$ are given by
\bea
f({\mathbf p})&=&g^{(0)}({\mathbf p})+i\frac{2\pi}{E_{\mathbf p}}
\bigg\{\;{\mathcal K}^{+-}({\mathbf p})
+\Big[ \;g^{(0)}({\mathbf p})+(g^{(0)}({\mathbf p}))^2\; \Big]
\Big[ \;{\mathcal K}^{++}({\mathbf p})
+{\mathcal K}^{+-}({\mathbf p})\Big]\;\bigg\}\;,\non
f'({\mathbf p})&=&i\frac{2\pi}{E_{\mathbf p}}
\bigg\{\;{\mathcal K}^{++}({\mathbf p})
+\Big[ \;g^{(0)}({\mathbf p})+(g^{(0)}({\mathbf p}))^2\; \Big]
\Big[ \;{\mathcal K}^{++}({\mathbf p})
+{\mathcal K}^{+-}({\mathbf p})\Big]\;\bigg\}\;. 
\eea
The appearance of $\delta (p^0+q^0)$ in Eq.~(\ref{delt1}) could,
in general, result in the time dependence of the full propagator.
However, if we assume that the initial time $t_0$ is in the remote past
($t_0=-\infty$), while $x^0$ and $y^0$ of the full propagator are
finite, we can consequently drop $\delta (p^0+q^0)$.
This is because, when going to coordinate space, this term would
generate a factor $e^{-ip^0(x^0+y^0-2t^0)}$ that vanishes
according to the Riemann theorem.
Under this assumption the full propagator in \eqrf{delt1} can be
finally written as
\bea
\Delta (p,q)&=&\bigg\{\;\Delta ^{(0)}_{in}(p)
-2\pi i\delta (p^2-m^2)
\Big[\; \lambda _1 + f({\mathbf p})
\lambda _2\;\Big]\;\bigg\}\;\delta ^4(p-q)\;,  
\label{delpq}
\eea
which has the standard form generally assumed in the literature.

In this section we have discussed the effect of the non-local
source kernel $K$ on the solution of the DSE in a simple free
scalar field model. In terms of this model we have derived the
full propagator and shown that  $K$ entering the homogeneous part
of the solution brings its time dependence and breaks the time
translational symmetry. If one assumes, however, that the initial
time is in the remote past, one then finds that the time
dependence can be neglected and thus the time translational
symmetry is restored. In this case  the effect of $K$ can be
collected to the homogeneous solution in such a way that $K$ only
corrects the distribution function. In equilibrium, this
distribution function is just the Bose-Einstein distribution. 
In the above free scalar field model that neglects 
interaction the equilibrium cannot be, 
however, reached. For a more complicated
cases such as QCD, the effect of $K$ on the solution is more
involved. Generally $K$ can be treated as a special kind of
boundary conditions \cite{calzetta}. Therefore, the kernel $K$ is
absent in the derivation of the transport equation that is
presented in the following sections.

The non-local source $K$ is defined as the matrix element of
the density matrix $\rho$ on the initial states $\phi_{1,2}$ 
\bea 
\langle \phi _1,0|\rho |\phi _2,0\rangle 
&=&\exp [iK(\phi _{\pm})]\;. 
\eea 
The kernel $K$ can be expanded functionally as follows:
\bea 
K(\phi _{\pm})&=& K+ \int d^4x K_a(x)\phi ^a(x) +\frac 12\int
d^4x d^4x' K_{ab}\phi ^a(x)\phi ^b(x') +\cdots 
\eea 
with $a,b=+,-$. In general  $K$ is a complex functional of the fields.
From \eqrf{delt1} and (\ref{delpq})  the coefficient ${\mathcal
K}$ of the quadratic term in the functional expansion of the
source $K$ should be purely imaginary to ensure that the  spectral
function is real. For a general form of the density matrix $\rho
\sim \exp (-\int d^3k \b _ka_{\mathbf{k}}
^{\dagger}a_{\mathbf{k}})$, one can indeed   check that  when
expanding the kernel $K$  up to the quadratic term  the
coefficients are  all imaginary [see Eq.\ (2.31)
\cite{calzetta}].

It is interesting to note that the situation here is
similar to the pinch singularity, which arises when the
time variation of the distribution function \cite{pinch}
is neglected. The possible connection between
the non-local source kernel and the
pinch singularity will be discussed elsewhere.

\section{Kinetic part of the transport equation}

In Section V we derived the DSE for a gluon plasma in the
CTP formalism. The resulting DSE summarized in
\eqrftw{sch-dy-c1}{sch-dy-c2} is a non-linear
integro-differential equation, which cannot be solved
without further approximations.
The essential approximation usually made in the literature
\cite{blaizot} is based on the two-scale nature
of high energy QCD. There are two typical scales
for a multiparton system: the quantum scale,
which characterizes quantum fluctuations or parton
self-interactions, and the statistical-kinetic scale,
which measures the range of interactions between quasi-particles.
These interactions may be described
in a semiclassical way if these two
scales are well separated, i.e. if the local density of
quasi-particles is smaller than a critical density where particles
begin to overlap. The above situation
is well suited to the case of ultra-relativistic heavy-ion
collisions. Shortly after two highly Lorentz-contracted nuclei
pass through each other, a very strong background field is formed,
followed by the production of very high energy partons.
Since this occurs very early and
at a very short space-time scale, it is
purely a quantum process, thus the quasi-particle-based
semiclassical or the kinetic description generally fail.
As time goes on, the dense system of partons undergoes
an expansion and the local density may fall down to a level
where the quantum and classical scales can be well separated.
Through multiple collisions, the parton system can
thermalize and then its bulk
properties can be described in terms of hydrodynamics.

With the above physical scenario in mind, let us assume that the
space-time is discretized into cells of a size chosen such
that the separation between the quantum and kinetic
scales \cite{calzetta,geiger} is optimized.
Then, the correlation between
different cells will be negligible.
The 2-point correlation will not vanish,
only when two space-time points lie in the same cell.
Consequently, in the multiparton system and in a given cell,
one can neglect spatial inhomogeneity of the local gluon and quark
densities. Within each cell, one may therefore describe the
short-distance quantum dynamics analogously, as in vacuum or in
a homogeneous medium. The inhomogeneity of the spatial parton
distribution associated with particle
collisions appears only when moving from cell to cell.

Let us define a mass scale $\mu$ as the separating point of the
quantum and the kinetic scale. This implies that one may characterize
the dynamical evolution of the parton system by a short-range
quantum scale $\lambda _{qua}\le\frac 1{\mu}$, and a long-range
kinetic scale $\lambda _{kin}\ge\frac 1{\mu}$. The low-momentum
collective excitations that may develop at the particular momentum
scale $g\mu$ are thus well separated from the typical hard gluon
momenta $k\ge \mu$, provided that $g\ll 1$. The effect
of the classical field $A$ on the hard quanta involves the
coupling $gA$ to the hard propagator; it is thus of the order of
the soft wavelength $\sim 1/(g\mu)$. Hence, we have the following
characteristic scales:
\bea &&y=x_1-x_2\sim \frac 1{\mu}\; ,\;\;\;
\partial_y=\frac 12(\partial_1-\partial_2)\sim \mu
\;,\nonumber\\
&&X=\frac 12(x_1+x_2)\sim \frac 1{g\mu}\; ,\;\;\;
\partial_X=\partial_1+\partial_2\sim g\mu
\;,\nonumber\\
&&gA(X)\sim g\mu\; ,\;\;\; gF[A(X)]\sim g^2\mu^2 \;,
\label{condt}
\eea
where $X$ is the central point and $y$ the difference of
two coordinates $x_1$ and $x_2$ in a 2-point GF.
We see that $X$ labels the macroscopic
kinetic motion whereas $y$ characterizes
the microscopic quantum distance.

In order to take advantage of the assumed separation of scales
we first express all 2-point GFs appearing in
Eqs.\ (\ref{sch-dy3}), (\ref{sch-dy4}) 
and (\ref{sch-dy5}), (\ref{sch-dy6})
in terms of the new variables $X$ and $y$.
Then, to derive the transport equation from the DSE, we perform a
gradient expansion of these GFs under the conditions (\ref{condt}).
From Eqs.\ (\ref{sch-dy-c1}) and (\ref{sch-dy-c2}) 
it is clear that we will deal with such an integral as
$I=\int d^4x'\Pi (x_1,x')G(x',x_2)$.
In terms of $X$, $y$ and $y'=x'-x_2$
coordinates, the integral and its Fourier transforms,
with respect to the relative distance $y$,
can be expressed in the gradient expansion as:
\bea
I &=& \int d^4 y'\Big[ \Pi (X,y-y')G(X,y')
+\frac{1}{2}\partial _X\Pi (X,y-y')y'G(X,y')\nonumber\\
&&-\frac{1}{2}(y-y')\Pi (X,y-y')\partial _XG(X,y') \Big]
\;,\non
F[I]&=&\int d^4y\,e^{iqy}\,I
=\Pi (X,q)G(X,q)+\frac i2 \Big[ \partial _q\Pi (X,q)
\cdot \partial _XG(X,q)\nonumber\\
&&-\partial _X\Pi (X,q)\cdot \partial _qG(X,q)\Big] \;.
\label{col-g-exp}
\eea
Similarly, the Fourier transforms for
$I_1=\Pi (x_1)G(x_1,x_2)$
and $I_2=G(x_1,x_2)\Pi (x_2)$ are
\bea
F[I_1]&=&\Pi (X)G(X,q)-\frac i2\partial _X\Pi (X)
\cdot \partial _qG(X,q)\;,\nonumber\\
F[I_2]&=&G(X,q)\Pi (X)+\frac i2\partial _qG(X,q)
\cdot \partial _X\Pi (X)\;.
\eea

The 2-point GFs in Eqs.\ (\ref{sch-dy3}), (\ref{sch-dy4}) and
(\ref{sch-dy5}), (\ref{sch-dy6}) are not gauge covariant.
In order to obtain a gauge-covariant transport equation,
we must use a gauge-covariant 2-point GF defined by
\bee
\label{tg-g}
\tilde{G}(X,y)=V(X,x_1)G(x_1,x_2)V(x_2,X)\;,
\eee
where $V(z_1,z_2)$ is a Wilson link with respect to the
classical background field given by
\bee
V(z_1,z_2)={\rm T}_{\rm P}
\exp \bigg( ig\int _{{\rm P};z_2}^{z_1} dz_{\mu}A_{\mu} \bigg)\;,
\eee
where the integral stands for a path integral from point $z_2$ to
$z_1$ and ${\rm T}_{\rm P}$ denotes the ordered product along the
path; note that the path here is defined in coordinate space.
The Wilson link $V(z_1,z_2)$ transforms as
\bee
V(z_1,z_2)\rightarrow U(z_1)V(z_1,z_2)U^{-1}(z_2)\;,
\eee
where $U(z)=\exp (ig\omega ^a(z)t_A^a)$
is the gauge transformation under which the GF
$G(x_1,x_2)$ transforms as
\bee
G(z_1,z_2)\rightarrow
U(z_1)G(z_1,z_2)U^{T}(z_2) =U(z_1)G(z_1,z_2)U^{-1}(z_2)
\;,
\eee
which involves transformations at two different space-time points.
However, the CGF $\tilde{G}(X,y)$ transforms as
\bee
\label{tg1}
\tilde{G}(X,y)\rightarrow U(X)\tilde{G}(X,y)U^{-1}(X) \;,
\eee
where only the transformation at a single point $X$ is relevant.
The gauge-covariant Wigner function is the Fourier transform of
$\tilde{G}(X,y)$ with respect to $y$:
\bee
\tilde{G}(X,q)=\int
d^4y\,\tilde{G}(X,y)e^{iqy}\;.
\eee
Obviously $\tilde{G}(X,q)$ transforms
in the same way as $\tilde{G}(X,y)$ according to \eqrf{tg1}.

In general, the integration path in the Wilson link may be of
arbitrary shape, provided that its two end points are fixed; the
gauge-covariant Wigner function is thus not uniquely defined. This
ambiguity can be removed by requiring that the Fourier spectrum
variable $q$ in the Wigner function $\tilde{G}(X,q)$ corresponds
to the kinetic momentum in the classical limit. This constraint is
fulfilled by the straight-line path \cite{elze86}. Therefore, in
all our future calculations we imply the straight-line path.
The link operator defined on the straight-line path
has the following properties:
\bea
&&V(z_1,z_2)V(z_2,z_3)=V(z_1,z_3)\;,\nonumber\\
&&V(z_1,z_2)V(z_2,z_1)=1\;,\nonumber\\
&&V^{\dagger}(z_1,z_2)=V^{-1}(z_1,z_2)=V(z_2,z_1)\;.
\eea
Obviously, the Wilson link is unitary since $A=A^{\dagger}$ and
forms a group. With the above properties, we can write the
inverse relation for \eqrf{tg-g} as
\bee
\label{g-tg}
G(x_1,x_2)=V(x_1,X)\tilde{G}(X,y)V(X,x_2)\;.
\eee

Consider a straight-line path from $z_2$ to $z_1$ described by the
equation $z(s)=z_2+(z_1-z_2)s$ with $s=[0,1]$. The variation of
the path characterized by small changes of their end points
$dz_1$ and $dz_2$ causes the following
variation of $V$ \cite{elze86}:
\bea
\delta V(z_1,z_2)&=&igA(z_1)dz_1V(z_1,z_2)
-igV(z_1,z_2)A(z_2)dz_2\nonumber\\
&&-ig\int _0^1 ds V\Big( z_1,z(s)\Big)
F_{\mu\nu}\Big( z(s)\Big) V\Big( z(s),z_2\Big)\nonumber\\
&&\cdot (z_1-z_2)_{\mu}
\Big[dz_2+(dz_1-dz_2)s\Big] _{\nu} \;.
\label{del-v}
\eea
Using the above equation, we obtain
\bea
\partial _{x_1\mu}V(x_1,X)
&=&-\frac 12igV(x_1,X)A_{\mu}(X)
+igA_{\mu}(x_1)V(x_1,X)\nonumber\\
&&-ig\frac 38y_{\nu}F_{\nu\mu}\Big( \frac 34x_1+\frac 14x_2\Big)
\nonumber\\
\partial _{x_1\mu}V(X,x_2)
&=&\frac 12igA_{\mu}(X)V(X,x_2) -ig\frac 18y_{\nu}F_{\nu\mu}
\big(\frac 14x_1+\frac 34x_2\big)\;.
\label{px1v1}
\eea
The first equation can be written also as
\bee
\label{dv1}
D_{x_1\mu}V(x_1,X)=-\frac 12igV(x_1,X)A_{\mu}(X)
-ig\frac 38y_{\nu}F_{\nu\mu} \Big(\frac 34x_1+\frac 14x_2\Big)\;,
\eee
where the l.h.s. is $O(g\mu)$ and the two terms on the r.h.s.
are $O(g\mu)$ and $O(g^2\mu)$ respectively.

Taking the Hermitian conjugate of \eqrf{px1v1} and
interchanging $x_1$ and $x_2$, we obtain:
\bea
\partial _{x_2\mu}V(X,x_2)
&=&\frac 12igA_{\mu}(X)V(X,x_2)-igV(X,x_2)A_{\mu}(x_2)
\nonumber\\
&&-ig\frac 38y_{\nu}F_{\nu\mu}
\Big(\frac 14x_1+\frac 34x_2\Big)
\;, \nonumber\\
\partial _{x_2\mu}V(x_1,X)
&=&-\frac 12igV(x_1,X)A_{\mu}(X) -ig\frac
18y_{\nu}F_{\nu\mu}\Big(\frac 34x_1+\frac 14x_2\Big)\;.
\label{dx2v}
\eea
From the above results we can also find that
\bea
D_{x_1\nu}(V_1\tilde{G}V_2) &=&\frac 12V_1(D_{X\nu}\tilde{G})V_2
+V_1(\partial _{y\nu}\tilde{G})V_2
-ig\frac 38y_{\lambda}F_{\lambda\nu}\tilde{G}V_2\nonumber\\
&&+\frac 12igV_1\tilde{G}A_{\nu}V_2 -ig\frac
18V_1\tilde{G}y_{\lambda}F_{\lambda\nu}\;, 
\label{dx1nu}
\eea
with the following compact notations: $V_{1}\equiv V(x_1,X)$,
$V_{2}\equiv V(X,x_2)$, $\tilde{G}\equiv \tilde{G}(X,y)$ and
$D_{x_1\nu}\equiv D_{\nu}[A(x_1)]$.
The terms on the r.h.s. of \eqrf{dx1nu} are of
orders $g\mu$, $\mu$, $g^2\mu$, $g\mu$ and $g^2\mu$, respectively.
We also make the approximation
$F_{\lambda\nu}(\frac 14x_1+\frac 34x_2)\approx F_{\lambda\nu}(X)$
in all terms involving $F_{\lambda\nu}$ in the above equation,
because the corrections are of $O(g^3\mu)$.

From \eqrf{dx1nu} and its Hermitian conjugate
$(V_1\tilde{G}V_2)D_{x_2\nu}^{\dagger}$
we can derive the gauge condition for the gauge-covariant GF.
The background gauge condition is given by
$D^{ab}_{\mu}[A(x)]Q_{\mu}^b(x)=0$,
thus the gauge condition for
the Green function with respect to $x_1$ reads
\bee
\label{gc1}
D^{ab}_{\mu}[A(x_1)]G_{\mu\nu}^{bc}(x_1,x_2)=0\;,
\eee
where $G$ is one of $G^>$, $G^<$ or $G^C$.
With respect to $x_2$, the gauge condition is
\bee
D^{ac}_{\nu}[A(x_2)]G_{\mu\nu}^{bc}(x_1,x_2)=0\;,
\eee
whose complex conjugate is given by
\bee
\Big[ \delta ^{ac}\partial _{2\nu}
+i g [(t^d_A)^{ac}]^*A_{\nu}^d(x_2) \Big]
G_{\mu\nu}^{bc}(x_1,x_2)=0\;,
\eee
where we have used the fact that
$A$ and $G$ are real.
The $SU(3)$ generators in the adjoint
representation are Hermitian
$(t^d_A)^{\dagger}=t^d_A$, therefore we have
\bea
&&\Big[ \delta ^{ac}\partial _{2\nu}
+ig(t^d_A)^{ca}A_{\nu}^d(x_2) \Big] G_{\mu\nu}^{bc}(x_1,x_2)
\nonumber\\
&&=G_{\mu\nu}^{bc}(x_1,x_2) D_{\nu}^{\dagger;ca}[A(x_2)]=0\;.
\label{gc2}
\eea
From Eqs.(\ref{g-tg}), (\ref{dx1nu}), 
(\ref{gc1}), and (\ref{gc2}) we find that
\bea
D_{x_1\mu}(V_1\tilde{G}_{\mu\nu}V_2) &=&
V_1\bigg\{ \frac 12 (D_{X\mu}\tilde{G}_{\mu\nu}) +(\partial
_{y\mu}\tilde{G}_{\mu\nu})
-ig\frac 38 y_{\lambda}F_{\lambda\mu}\tilde{G}_{\mu\nu}
\nonumber\\
&&+\frac 12ig\tilde{G}_{\mu\nu}A_{\mu}
-ig\frac 18\tilde{G}_{\mu\nu}y_{\lambda}F_{\lambda\mu}
\bigg\} V_2=0 \;,
\eea
and
\bea
(V_1\tilde{G}_{\nu\mu}V_2)D_{x_2\mu}^{\dagger}
&=&V_1\bigg\{ \frac 12(\tilde{G}_{\nu\mu}D^{\dagger}_{X\mu})
-(\partial _{y\mu}\tilde{G}_{\nu\mu})
-ig\frac 38\tilde{G}_{\nu\mu}y_{\lambda}F_{\lambda\mu}
\nonumber\\
&&-\frac 12igA_{\mu}\tilde{G}_{\nu\mu} -ig\frac
18y_{\lambda}F_{\lambda\mu}\tilde{G}_{\nu\mu} \bigg\} V_2=0 \;,
\eea
where we kept only terms up to $O(g^2\mu)$. Taking the sum and the
difference of the above two equations, we derive the gauge
condition for the gauge-covariant $\tilde{G}$:
\bea
\frac 12\partial _{X\mu}\tilde{G}_{\{\mu\nu\}} +\partial
_{y\mu}\tilde{G}_{[\mu\nu]} +\frac 12 i g \Big[
\tilde{G}_{\{\mu\nu\}}, A_{\mu} \Big] =0
\;,\nonumber\\
\frac 12\partial _{X\mu}\tilde{G}_{[\mu\nu]} +\partial
_{y\mu}\tilde{G}_{\{\mu\nu\}} +\frac 12 i g \Big[
\tilde{G}_{[\mu\nu]}, A_{\mu}\Big] =0 \;,
\eea
where $\tilde{G}_{\{\mu\nu\}}$ and $\tilde{G}_{[\mu\nu]}$
are defined by $\tilde{G}_{\mu\nu}+\tilde{G}_{\nu\mu}$ and
$\tilde{G}_{\mu\nu}-\tilde{G}_{\nu\mu}$ respectively.
For simplicity we also neglect the terms of
order higher than $O(g\mu)$. 

If we assume that $\tilde{G}$ is symmetric in its Lorentz
indices, then up to $O(g\mu) $ we obtain
\bee
\label{ggcond}
\partial _{X\mu}\tilde{G}_{\mu\nu}
+ig[\tilde{G}_{\mu\nu},A_{\mu}]=0\;,\;\;
\partial _{y\mu}\tilde{G}_{\mu\nu}=0 \;.
\eee
The second equation tells us that $\tilde{G}_{\mu\nu}$ is
transversal up to $O(g\mu)$.

Taking the second covariant derivative of \eqrf{dx1nu}, we
obtain the covariant d'Alembertian operator:
\bea
D^2_{x_1}(V_1\tilde{G}V_2) &=&\frac 14
V_1(D^2_{X}\tilde{G})V_2 +V_1(\partial _{y}\cdot
D_X\tilde{G})V_2 +V_1(\partial ^2_{y}\tilde{G})V_2
\nonumber \\
&&-ig\frac 38y_{\lambda}F_{\lambda\nu}(D_{X\nu}\tilde{G})V_2
-ig\frac 34y_{\lambda}F_{\lambda\nu}
(\partial _{y\nu}\tilde{G})V_2
\nonumber \\
&&+ig\frac 12V_1(D_{X\nu}\tilde{G})A_{\nu}V_2
+ig\frac 14V_1\tilde{G}(\partial _{X\nu}A_{\nu})V_2
\nonumber \\
&&-ig\frac 18V_1(D_{X\nu}\tilde{G})y_{\lambda}F_{\lambda\nu}
-ig\frac 14V_1(\partial _{y\nu}\tilde{G})
y_{\lambda}F_{\lambda\nu}
\nonumber \\
&&+igV_1(\partial _{y\nu}\tilde{G})A_{\nu}V_2
-\frac 14g^2V_1\tilde{G}A^2V_2 \;,
\label{du1}
\eea
where we kept only terms up to $O(g^3\mu ^2)$, neglecting all
higher order ones. The resulting equation for
$(V_1\tilde{G}V_2)D^{\dagger 2}_{x_2}$ can be derived by taking
the Hermitian conjugate of $D^2_{x_1}(V_1\tilde{G}V_2)$ and then
interchanging $x_1$ with $x_2$:
\bea
(V_1\tilde{G}V_2)D^{\dagger 2}_{x_2} &=&
\frac 14V_1(\tilde{G}D^{\dagger 2}_{X})V_2
-V_1(\tilde{G}\stackrel{\leftarrow}{\partial }_{y} \cdot
D^{\dagger}_{X})V_2
+V_1(\partial ^2_{y}\tilde{G})V_2\nonumber \\
&&-ig\frac 38V_1(\tilde{G}D^{\dagger}_{X\nu})y_{\lambda}F_{\lambda\nu}
+ig\frac 34V_1(\partial _{y\nu}\tilde{G})
y_{\lambda}F_{\lambda\nu}\nonumber \\
&&-ig\frac 12V_1A_{\nu}(\tilde{G}D^{\dagger}_{X\nu})V_2
-ig\frac 14V_1(\partial _{X\nu}A_{\nu})\tilde{G}V_2
\nonumber \\
&&-ig\frac 18y_{\lambda}F_{\lambda\nu}(\tilde{G}D^{\dagger}_{X\nu})V_2
+ig\frac 14y_{\lambda}F_{\lambda\nu}(\partial _{y\nu}\tilde{G})V_2
\nonumber \\
&&+igV_1A_{\nu}(\partial _{y\nu}\tilde{G})V_2
-\frac 14g^2V_1A^2\tilde{G}V_2 \;.
\label{du2}
\eea
We take the difference between \eqrf{du1} and \eqrf{du2}:
\bee
\label{v1dv2}
V_1\Delta V_2\equiv
D^2_{x_1}(V_1\tilde{G}V_2)
-(V_1\tilde{G}V_2)D^{\dagger 2}_{x_2} \;,
\eee
where $\Delta $ is defined by
\bea
\Delta &=&\partial _{y}\cdot D_X\tilde{G}
+\tilde{G}\stackrel{\leftarrow}{\partial }_{y}\cdot D^{\dagger}_{X}
-ig\frac 38y_{\lambda}F_{\lambda\nu}(D_{X\nu}\tilde{G})
+ig\frac 38(\tilde{G}D^{\dagger}_{X\nu})
y_{\lambda}F_{\lambda\nu}
\nonumber\\
&&-igy_{\lambda}F_{\lambda\nu}(\partial _{y\nu}\tilde{G})
-ig(\partial _{y\nu}\tilde{G})y_{\lambda}F_{\lambda\nu}
\nonumber\\
&&-ig\frac 18(D_{X\nu}\tilde{G})y_{\lambda}F_{\lambda\nu}
+ig\frac 18y_{\lambda}F_{\lambda\nu}(\tilde{G}
D^{\dagger}_{X\nu})\nonumber\\
&&+ig(\partial _{y\nu}\tilde{G})A_{\nu}
-igA_{\nu}(\partial _{y\nu}\tilde{G}) \;.
\eea
Keeping terms up to $O(g^2\mu ^2)$, we obtain
\bea
\Delta &=&\partial _{y}\cdot D_X\tilde{G}
+\tilde{G}\stackrel{\leftarrow}{\partial }_{y}
\cdot D^{\dagger}_{X}
-igy_{\lambda}F_{\lambda\nu}(\partial _{y\nu}\tilde{G})
\non
&&-ig(\partial _{y\nu}\tilde{G})y_{\lambda}F_{\lambda\nu}
+ig(\partial _{y\nu}\tilde{G})A_{\nu}
-igA_{\nu}(\partial _{y\nu}\tilde{G})\;.
\eea
Taking the difference of the l.h.s. of Eqs.\ (\ref{sch-dy-c1}) and
(\ref{sch-dy-c2}), and using Eqs.\ (\ref{du1}) 
and (\ref{du2}) we get
\bea
\Delta '&=&\partial _{y}\cdot
D_X\tilde{G}_{\alpha\gamma}^{C} +\tilde{G}_{\alpha\gamma}^{C}
\stackrel{\leftarrow}{\partial }_{y}\cdot D^{\dagger}_{X}
+igy_{\lambda}F_{\nu\lambda}(\partial _{y\nu}
\tilde{G}_{\alpha\gamma}^{C}) +ig(\partial
_{y\nu}\tilde{G}_{\alpha\gamma}^{C})
y_{\lambda}F_{\nu\lambda}\nonumber\\
&&+ig(\partial _{y\nu}\tilde{G}_{\alpha\gamma}^{C})A_{\nu}
-igA_{\nu}(\partial _{y\nu}\tilde{G}_{\alpha\gamma}^{C})
-2ig(F_{\alpha\beta}\tilde{G}_{\beta\gamma}^{C}
-\tilde{G}_{\alpha\beta}^{C}F_{\beta\gamma}) \;,
\label{trkin1}
\eea
where we restored the Lorentz index for $\tilde{G}^C(X,y)$
and suppressed two Wilson links in front of and behind the
above expression. With respect to the
relative coordinate $y$, the Fourier transform reads
\bea
F[\Delta ']&=& -i \Big\{ q\cdot
D_X\tilde{G}_{\alpha\gamma}^{C} +\tilde{G}_{\alpha\gamma}^{C}
q\cdot D^{\dagger}_{X} +gq_{\nu}F_{\nu\lambda}(\partial
_{q\lambda} \tilde{G}_{\alpha\gamma}^{C}) +g(\partial
_{q\lambda}\tilde{G}_{\alpha\gamma}^{C})
q_{\nu}F_{\nu\lambda}\nonumber\\
&&+ig(\tilde{G}_{\alpha\gamma}^{C}q\cdot A
-q\cdot A\tilde{G}_{\alpha\gamma}^{C})
+2g(F_{\alpha\beta}\tilde{G}_{\beta\gamma}^{C}
-\tilde{G}_{\alpha\beta}^{C}F_{\beta\gamma})\Big\} \;,
\label{trkin2}
\eea
where $\tilde{G}^{C}\equiv \tilde{G}^{C}(X,q)$,
$A\equiv A(X)$ and $F\equiv F(X)$.

Equations (\ref{trkin1}) and (\ref{trkin2}) just describe the
kinetic part of the transport equation.
The collision part will be derived in the next section.
Neglecting the collision terms,
which are of order at least $g^4\mu ^2$,
we have the kinetic equation in the following compact form:
\bea
&& q\cdot
\partial _X\tilde{G}_{\alpha\gamma}^{C}
+ig(\tilde{G}_{\alpha\gamma}^{C}q\cdot A -q\cdot
A\tilde{G}_{\alpha\gamma}^{C}) \non &&+\frac 12
gq_{\nu}F_{\nu\lambda}(\partial _{q\lambda}
\tilde{G}_{\alpha\gamma}^{C}) +\frac 12 g(\partial
_{q\lambda}\tilde{G}_{\alpha\gamma}^{C}) q_{\nu}F_{\nu\lambda}\non
&&+g(F_{\alpha\beta}\tilde{G}_{\beta\gamma}^{C}
-\tilde{G}_{\alpha\beta}^{C}F_{\beta\gamma})=0\;. 
\label{treq1}
\eea
The above equation is located at the collective coordinate
$X$ and is {\it gauge covariant} under the local gauge
transformation $U(X)$, i.e. it transforms as $U(\cdots )U^{-1}$.
Indeed, noting that both $F_{\m\n}$ and
$\tilde{G}_{\alpha\gamma}^{C}$ are gauge covariant and
$\partial_{q\m}$ does not affect $U$, it is obvious that the last
three terms are gauge covariant. To verify that the first two
terms are also preserving gauge covariance, we explicitly write
down their transformations: 
\bea q\cdot \partial
_X\tilde{G}_{\alpha\gamma}^{C} &\ra & U(q\cdot \partial_X
\tilde{G}_{\alpha\gamma}^{C})U^{-1} +(q\cdot \partial_X
U)\tilde{G}_{\alpha\gamma}^{C}U^{-1} \non
&&+U\tilde{G}_{\alpha\gamma}^{C}q\cdot \partial_X U^{-1}\;, \non
ig\tilde{G}_{\alpha\gamma}^{C}q\cdot A &\ra &
igU\tilde{G}_{\alpha\gamma}^{C}q\cdot A\; U^{-1}
-U\tilde{G}_{\alpha\gamma}^{C}q\cdot \partial_X U^{-1}\;, \non -ig
q\cdot A \tilde{G}_{\alpha\gamma}^{C} &\ra & -ig U q\cdot A
\tilde{G}_{\alpha\gamma}^{C} U^{-1} \non && - ( q\cdot \partial_X
U ) \tilde{G}_{\alpha\gamma}^{C} U^{-1} \;, 
\nonumber
\eea
from which it is clearly seen that the
sum of the first two terms in \eqrf{treq1} indeed transforms as
$U(\cdots )U^{-1}$ and therefore preserves the gauge covariance.

The quantum kinetic equation (\ref{treq1}) was derived in a
quite general and transparent manner in the context of BG-QCD and
CTP formalism. No further approximations or requirements going
beyond gradient expansion were used to obtain Eq.~(\ref{treq1}).
We note that a result similar to
Eq.\ (\ref{treq1}) was previously obtained
in Ref.\ \cite{mrow}.
There, however, based on Ref.~\cite{elze88}, the transport
equation was derived by making the
gradient expansion of the equation of motion for the Wigner
function (not in CTP formalism). Additionally the author of 
Ref.\ \cite{mrow} made the derivation in the fundamental color space
(not the adjoint space) in QCD (not in BG-QCD). 
Finally, in Ref.\ \cite{mrow}, 
\eqrf{treq1} was obtained by 
assuming that the Wigner function is proportional to the
quadratic product of the generators of the
fundamental representation.
Our approach is quite general and does not require any
specific assumptions on the structure of the Wigner function.

In the following we show that \eqrf{treq1} is a natural quantum
generalization of the classical Boltzmann equation. In particular
the colour charge precession will be explicitly identified in the
quantum description of the
colour charge kinetics given by \eqrf{treq1}.

The classical kinetic equation for the colour singlet distribution
function $f(x,p,Q)$ is \cite{wong,heinz,elze86,jalilian,kelly,litim}:
\bea
\label{wongeq}
&&p_{\mu}[\partial
_{\mu}-gQ^aF_{\mu\nu}^a
\partial _{p\nu} \non
&&-gf^{abc}A_{\mu}^b(x)Q^c\partial _{Q^a}] f(x,p,Q)=0\;,
\eea
where $Q^a$ is the classical colour charge 
and $a=1,..., N^2_c-1$.

Comparing \eqrf{wongeq} with the quantum expression (\ref{treq1}),
it is clear that the colour singlet distribution function $f$ is
replaced by the gauge covariant Wigner function $\tilde{G}^C$, 
which is a colour matrix in the adjoint representation. One can
also recognize that the first, third and fourth terms of
\eqrf{treq1} are the quantum generalization of the first two terms
in \eqrf{wongeq}. The last term in \eqrf{treq1} appears from the
covariant operators and hence it does not appear in the classical
equation. This term can be written in different form by using
generators of Lorentz transformation in {\em vector}
representation. A similar term can be found for the quark, but
expressed through generators in {\em spinor} representation.


Particularly interesting is the appearance of the second term in
\eqrf{treq1}. We have seen that its presence is crucial to assure
the gauge covariance of the Vlasov equation. 
This term has an interesting physical meaning. It is the quantum
analogue to the color charge precession in the classical kinetic
equation. To see this more clearly, one can expand
$\tilde{G}^C_{\a\b}(X,q)$ with respect to the expansion 
parameter $gT^a A^a_{\m}(X)$ from the Wilson link in \eqrf{tg-g}. 
This expansion can be also understood as the result of 
the $AQQ$, $AQQQ$ and $AAQQ$ vertices. Then we have 
\bea
\tilde{G}^C_{\a\b}(X,q)&=&N_{0\a\b}(X,q)+gT^a A^a_{\m}(X) 
N_{1\a\b;\m}(X,q) \non
&&+g^2 T^a T^b A^a_{\m}(X) A^b_{\n}(X) N_{2\a\b;\m\n}(X,q)\non 
&&+\cdots \;, 
\label{expansion} 
\eea
where $T^a$ are quantum analogues to the classical color charges
$Q^a$; $N_{i\a\b}(X,q)$ with $i=0,1,2,\cdots $ 
are color singlet functions. Each term of 
the expansion corresponds to an order of the color 
inhomogeneity in the gluonic medium due to its interaction 
with the background field. If the background field is 
refered to the soft mean field, its magnitude 
should vanish when the system approaches equilibrium. 
Then only the first singlet term survives in \eqrf{expansion}, 
which means the color homogeneity of the gluonic medium. 
This is somewhat similar to the 
multipole expansion for an electromagnetic source 
where the moments of dipole, quadrupole etc. 
describe increasing orders of spatial 
inhomogeneity for the electromagnetic charges. 
In weak coupling, as the lowest order approximation, 
we keep only the first two terms in \eqrf{expansion}. 
Then the second term of \eqrf{treq1} becomes
\begin{equation}
ig(\tilde{G}_{\alpha\gamma}^{C}q\cdot A -q\cdot
A\tilde{G}_{\alpha\gamma}^{C})\simeq -g f^{abc} q_{\m} A^b_{\m}
T^c \partial_{T^a} \tilde{G}_{\alpha\gamma}^{C}
\end{equation}
which reproduces the classical
color precession term in Eq.\ (\ref{wongeq}).

The covariant derivative $D_X\sim gA(X)\sim g\m$ 
and therefore first two terms of \eqrf{treq1}
are at leading order, $O(g\m ^2)$, 
while other terms are at subleading
order, $O(g^2\m ^2)$. In the vicinity of equilibrium the natural
scale in the system is the temperature $T$. The mean distance
between particles is of the order of $\sim 1/T$, while $1/(gT)$
characterizes the scale of collective excitations
\cite{blaizot,blaizot1}. For small coupling constant $g$ these two
scales are well separated. The covariant Wigner functions can be
expanded around their equilibrium values:
$\tilde{G}=\tilde{G}^{(0)}+\delta\tilde{G}$, where the equilibrium
function $\tilde{G}^{(0)}$ is a colour singlet and the fluctuation
$\delta\tilde{G}\sim g^2\tilde{G}$. Typical scales are $q\sim T$,
$D_X\sim g^2T$, $gF\sim (D_X)^2\sim g^4T^2$. Thus at leading
order, only the first term of \eqrf{treq1} survives and the
precession term vanishes because of the color-singlet nature of
$\tilde{G}^{(0)}$. 
The linearized version of Eq.\ (\ref{treq1}) with 
respect to $\delta \tilde G$
corresponds to the equation formulated in 
the background Coulomb gauge in Ref.~\cite{blaizot}.

The quantum fluctuations near equilibrium were also
considered in Ref.\ \cite{litim1} in the context of 
the classical collisionless transport equation. 
It is quite natural to carry out the same study 
from our quantum approach. First, BG-QCD deals 
with the classical field and the quantum fluctuation 
in a systematic way. The quantum field plays the 
similar role to the field fluctuation in Ref.\ \cite{litim1}. 
Second, in the quantum approach, 
corresponding to the phase-space distribution, 
we deal with the GFs which can 
be expanded around their equilibrium values 
following Eq.\ (\ref{expansion}). 
One also needs to complete 
the equations by including the field
equation (\ref{eqm}) where the averaged induced current 
is related to the 2- and 3-point GFs 
which finally depend on $\tilde{G}$. 
Thus it can be expanded according to 
Eq.\ (\ref{expansion}) as well.

The analogy and differences of the quantum and classical
Boltzmann equations can also be clearly exhibited when formulating
the equations for the colour moments. Corresponding to
\eqrf{treq1} one gets
\bea
&& q\cdot \partial _X h_{\alpha\gamma}
+g q_{\nu} F_{\nu\lambda}^a \partial _{q\lambda}
h_{\alpha\gamma}^a
\non
&& +g ( F_{\alpha\beta}^ah_{\beta\gamma}^a
-h_{\alpha\beta}^aF_{\beta\gamma}^a )=0\;,
\label{treq01} \\
&& q\cdot \partial _X h_{\alpha\gamma}^a
+gf^{abc}q\cdot A^bh_{\alpha\gamma}^c
\non
&& +g q_{\nu} F_{\nu\lambda}^b\partial _{q\lambda}
\frac 12[h_{\alpha\gamma}^{ab}+h_{\alpha\gamma}^{ba}]
\non
&& +g(F_{\alpha\beta}^bh_{\beta\gamma}^{ab}
-h_{\alpha\beta}^{ba}F_{\beta\gamma}^b)=0\;,
\label{treq2}
\eea
where we define: $h_{\alpha\gamma}={\rm
Tr}(\tilde{G}_{\alpha\gamma}^{C})$, $h_{\alpha\gamma}^a={\rm
Tr}(T^a\tilde{G}_{\alpha\gamma}^{C})$ and
$h_{\alpha\gamma}^{ab}={\rm
Tr}(T^aT^b\tilde{G}_{\alpha\gamma}^{C})$.

The classical equations for colour moments of
$f(x,p,Q)$ are given by
\bea
&&p\cdot\partial _{x}f(x,p)-g p_{\mu}F_{\mu\nu}^a
\partial _{p\nu}f^a(x,p)=0\;,
\label{wong01}\\
&&p\cdot\partial _{x}f^a(x,p)+gf^{abc}p\cdot A^b(x)f^c(x,p)
\non
&&-g p_{\mu}F_{\mu\nu}^b\partial _{p\nu}f^{ab}(x,p)=0\;,
\label{wong1}
\eea
where $f^{ab}(x,p)=\int dQ Q^a Q^b f(x,p,Q)$,  $f(x,p)=\int dQ
f(x,p,Q)$ and $f^a(x,p)=\int dQ Q^a f(x,p,Q)$.

Comparing Eq.\ (\ref{treq01}) with (\ref{wong01}) and
Eq.\ (\ref{treq2}) with (\ref{wong1}) one sees that, 
apart from terms like 
$(F_{\alpha\beta}h_{\beta\gamma}-h_{\alpha\beta}F_{\beta\gamma})$,
which come from the covariant operators, the quantum and
classical equations have a similar structure. The identification of
the colour precession term in Eq.\ (\ref{treq1}) is straightforward.

\section{Collision Part}

In this section we will derive the gauge covariant collision
part of the transport equation in a pure gluon plasma. The
collision part is of the higher order in $g$ as it is suppressed
at least by $g^2$ relative to the kinetic part.

The collision part is derived by taking the difference of the
SE terms in the r.h.s. of Eqs. (\ref{sch-dy-c1}) and
(\ref{sch-dy-c2}) and then performing a gradient expansion.
First, these equations should be expressed in terms of the
gauge covariant Wigner functions $\tilde{G}$ and $\tilde{\Pi }$.

The integral of \eqrf{col-g-exp} can be written as
\bea
I&=&\int d^4y'V\Big( x_1,X+\frac{y'}{2}\Big) \tilde{\Pi}\Big(
X+\frac{y'}{2},y-y'\Big) V\Big(
X+\frac{y'}{2},X-\frac{y-y'}{2}\Big)
\nonumber\\
&&\times\tilde{G}\Big( X-\frac{y-y'}{2},y'\Big)
V\Big( X -\frac{y-y'}{2},x_2\Big)
\nonumber\\
&\approx &V(x_1,X)\bigg\{ \int d^4y'
\Big[ \tilde{\Pi}\Big( X+\frac{y'}{2},y-y'\Big)
\tilde{G}\Big( X -\frac{y-y'}{2},y'\Big) \nonumber\\
&&-ig\frac{y'}{2}\cdot A(X)
\tilde{\Pi}(X,y-y') \tilde{G}(X,y')\nonumber\\
&&-\tilde{\Pi}(X,y-y') \tilde{G}(X,y')
ig\frac{y-y'}{2}\cdot A(X)\nonumber\\
&&+\tilde{\Pi}(X,y-y')
ig\frac{y}{2}\cdot A(X)
\tilde{G}(X,y') \Big] \bigg\}V(X,x_2)\nonumber\\
&&+{\rm higher\,\,order\,\,in} \,\,y\,\,{\rm or}\,\, y' \;,
\label{col1}
\eea
where we have made a gradient expansion of the Wilson link
operators and kept only leading terms in  $y$ or $y'$,
applying \eqrf{del-v}:
\bea
V\Big( x_1,X+\frac{y'}{2}\Big) &\approx
&V(x_1,X)\Big[ 1-ig\frac{y'}{2}\cdot A(X)
+\cdots \Big] \nonumber\\
V\Big( X-\frac{y-y'}{2},x_2 \Big)
&\approx &\Big[ 1-ig\frac{y-y'}{2}\cdot A(X)
+\cdots \Big] V(X,x_2)\;,\nonumber\\
V\Big( X+\frac{y'}{2},X-\frac{y-y'}{2} \Big)
&\approx &1+ig\frac{y}{2}\cdot A(X) \;.
\eea
The Fourier transform of the term inside the curly bracket
$\{\cdots \}$ in \eqrf{col1} reads
\bea
F[I]&=&\tilde{\Pi}(X,q)\tilde{G}(X,q)+\frac{i}{2} \Big\{
[\partial _q\tilde{\Pi}(X,q)]\cdot
[\tilde{G}(X,q)D_X^{\dagger}]\nonumber\\
&&+[\partial _q\tilde{\Pi}(X,q)]
\cdot [ D_X\tilde{G}(X,q)]
-[D_X\tilde{\Pi}(X,q)]\cdot [\partial _q\tilde{G}(X,q)]
\nonumber\\
&&-[\tilde{\Pi}(X,q)D_X^{\dagger}]
\cdot [\partial _q\tilde{G}(X,q)]\Big\}
+\frac{1}{2}g\partial _q [\tilde{\Pi}(X,q)
A(X)\tilde{G}(X,q)]
\nonumber\\
&\equiv&\tilde{\Pi}\tilde{G}+\frac{i}{2}
\Big[ \partial _q\tilde{\Pi}\cdot\tilde{G}D_X^{\dagger}
+\partial _q\tilde{\Pi}\cdot D_X\tilde{G}
-D_X\tilde{\Pi}\cdot\partial _q\tilde{G}
-\tilde{\Pi}D_X^{\dagger}\cdot\partial _q\tilde{G}\Big]
\nonumber\\
&&+\frac{1}{2}g\partial _q(\tilde{\Pi}A\tilde{G})\;,\label{fi}
\eea
where in the second equality we suppressed the arguments of
$\tilde{G}(X,q)$, $\tilde{\Pi}(X,q)$ and $A(X)$. In the
transformation from the coordinate to the momentum space
we use the following replacement:
$\partial _{y}\rightarrow -iq$ and
$y\rightarrow -i\partial _{q}$.

In the above equations all terms containing the derivatives 
$\partial _q$ and $\partial _X$ are suppressed by $g$ relative to
the leading term $\tilde{\Pi}\tilde{G}$. Keeping only the
lowest order contribution to Eq.~(\ref{fi}),
the collision term reads
\bee \label{tr2} I_{coll}=\tilde{G}^R\tilde{\Pi
}^C+\tilde{G}^C\tilde{\Pi }^A -\tilde{\Pi
}^C\tilde{G}^A-\tilde{\Pi }^R\tilde{G}^C \;,
\eee
where we drop the Wilson links
$V(x_1,X)$ and $V(X,x_2)$ as they can be
cancelled with those in the kinetic part.
To further simplify the
collisions term, we use the following relations:
\bee \bigg\{
\ba{l}\tilde{G}^R=\frac 12(\tilde{G}^A+\tilde{G}^R) +\frac
12(\tilde{G}^>-\tilde{G}^<)
\\\tilde{G}^A=\frac 12(\tilde{G}^A+\tilde{G}^R)
-\frac 12(\tilde{G}^>-\tilde{G}^<)\ea ,\,\,\,\,
\bigg\{\ba{l}\tilde{\Pi }^R=\frac 12(\tilde{\Pi}^A+\tilde{\Pi}^R)
+\frac 12(\tilde{\Pi}^<-\tilde{\Pi}^>)
\\\Pi ^A=\frac 12(\tilde{\Pi}^A+\tilde{\Pi}^R)
-\frac 12(\tilde{\Pi}^<-\tilde{\Pi}^>)\ea \,
 \eee
to obtain
\bea
I_{coll}&=&\tilde{G}^R\tilde{\Pi }^C
+\tilde{G}^C\tilde{\Pi }^A
-\tilde{\Pi }^C\tilde{G}^A-\tilde{\Pi }^R\tilde{G}^C
\nonumber\\
&=&\{\tilde{G}^<,\tilde{\Pi }^>\}
-\{\tilde{G}^>, \tilde{\Pi }^<\}
+\frac 12 \Big[\tilde{G}^>
+\tilde{G}^<, \tilde{\Pi}^A+\tilde{\Pi}^R\Big]
+\frac 12\Big[\tilde{\Pi }^>
+\tilde{\Pi }^<, \tilde{G}^A
+\tilde{G}^R\Big] \;,
\label{col2}
\eea
where $\tilde{G}$ and $\tilde{\Pi}$ are the matrices in the
colour and Lorentz indices.

While the transport equation
is derived by taking the difference of 
Eqs.\ (\ref{sch-dy-c1}) and (\ref{sch-dy-c2}),
the sum of the two equations gives the mass-shell equation:
\bea
&&\left( q^2-\frac 14\partial _X^2\right)\tilde{G}
+\frac 14 ig\Big[ (\partial _X\cdot A),\tilde{G}\Big]
+\frac 14 g^2 \Big\{ A\cdot A,\tilde{G}\Big\}\non
&&-\frac 14 g^2 A_{\nu}\tilde{G}A_{\nu}
-\frac 12 ig \Big[ (\partial _{X\nu}\tilde{G}),A_{\nu}\Big]
+\frac 14 ig q_{\nu}
\Big[ F_{\lambda ,\nu}, (\partial _{q\lambda}\tilde{G})\Big]
+ig \Big\{ F,\tilde{G} \Big\} \non
&&=\frac 12 \Big( \tilde{G}^R\tilde{\Pi }^C
+\tilde{G}^C\tilde{\Pi }^A
+\tilde{\Pi }^C\tilde{G}^A
+\tilde{\Pi }^R\tilde{G}^C \Big)\;.
\label{mass}
\eea
This equation provides the constraint on
the gauge covariant GF.

The collision term (\ref{tr2}), (\ref{col2}) combined with the
kinetic part (\ref{treq1}) gives a complete result for the
transport equation for a pure gluon plasma. A possible way of 
obtaining some physical insight 
and interpretation of this equation is to
consider the simple case that the system is near equilibrium. In
this case, we decompose the GF $\tilde{G}$ and the SE tensor
$\tilde{\Pi}$ as: $\tilde{G}=\tilde{G}^{(0)}+\delta\tilde{G}$ and
$\tilde{\Pi}=\tilde{\Pi}^{(0)}+\delta\tilde{\Pi}$, where their
equilibrium values $\tilde{G}^{(0)}$ and $\tilde{\Pi}^{(0)}$ are
colour singlets and $\delta\tilde{G}$ and $\delta\tilde{\Pi}$
denote deviations from equilibrium. Obviously, both the
collision and kinetic parts vanish for $\tilde{G}^{(0)}$ and
$\tilde{\Pi}^{(0)}$.

In the vicinity of equilibrium the natural scale is the
temperature $T$. The mean distance between particles is of the
order of $1/T$, whereas $1/(gT)$ characterizes the scale of the
collective excitations \cite{blaizot,blaizot1}. In the weak
coupling limit for $g\ll 1$ these two scales are well separated.
Therefore we have: $q\sim T$, $D_X\sim g^2T$ and $gF\sim
(D_X)^2\sim g^4T^2$ and the fluctuations $\delta\tilde{G}\sim
g^2\tilde{G}^{(0)}$ and $\delta\tilde{\Pi}\sim
g^2\tilde{\Pi}^{(0)}$. In the leading order the Boltzmann equation
reads
\bea
&&q\cdot \partial _X\delta\tilde{G}^C_{\alpha\gamma}
-gF_{\lambda\nu}q_{\nu}\partial _{q\lambda}
\tilde{G}^{(0)C}_{\alpha\gamma}
+g(F_{\alpha\beta}\tilde{G}^{(0)C}_{\beta\gamma}
-\tilde{G}_{\alpha\beta}^{(0)C}F_{\beta\gamma})
\nonumber\\
&&= i \frac 12\Big[ \tilde{G}^{(0)<}_{\alpha\beta}
\delta\tilde{\Pi}^{>}_{\beta\gamma}
+\delta\tilde{\Pi}^{>}_{\alpha\beta}
\tilde{G}^{(0)<}_{\beta\gamma}
-\tilde{G}^{(0)>}_{\alpha\beta}
\delta\tilde{\Pi}^{<}_{\beta\gamma}
-\delta\tilde{\Pi}^{<}_{\alpha\beta}
\tilde{G}^{(0)>}_{\beta\gamma}
\nonumber \\
&&+\delta \tilde{G}^{<}_{\alpha\beta}
\tilde{\Pi}^{(0)>}_{\beta\gamma} +\tilde{\Pi}^{(0)>}_{\alpha\beta}
\delta \tilde{G}^{<}_{\beta\gamma} -\delta
\tilde{G}^{>}_{\alpha\beta} \tilde{\Pi}^{(0)<}_{\beta\gamma}
-\tilde{\Pi}^{(0)<}_{\alpha\beta} \delta
\tilde{G}^{>}_{\beta\gamma}\Big]\;,
\label{trtiny}
\eea
where the color commutator $[ \tilde{G}^{(0)C},q\cdot A ]$ has
been neglected, since $\tilde{G}^{(0)}$ is a colour singlet. In
deriving the above equation we have also used the
following approximations:
\bea
(\tilde{G}^{(0)>}_{\alpha\beta} +\tilde{G}^{(0)<}_{\alpha\beta})
(\delta\tilde{\Pi}^A_{\beta\gamma}
+\delta\tilde{\Pi}^R_{\beta\gamma}) & = &
(\delta\tilde{\Pi}^A_{\alpha\beta}
+\delta\tilde{\Pi}^R_{\alpha\beta})
(\tilde{G}^{(0)>}_{\beta\gamma} +\tilde{G}^{(0)<}_{\beta\gamma})
\;,\nonumber\\
(\delta\tilde{G}^{>}_{\alpha\beta}
+\delta\tilde{G}^{<}_{\alpha\beta})
(\tilde{\Pi}^{(0)A}_{\beta\gamma}
+\tilde{\Pi}^{(0)R}_{\beta\gamma}) & = &
(\tilde{\Pi}^{(0)A}_{\alpha\beta}
+\tilde{\Pi}^{(0)R}_{\alpha\beta})
(\delta\tilde{G}^{>}_{\beta\gamma}
+\delta\tilde{G}^{<}_{\beta\gamma})\;.
\label{com}
\eea
We recall that up to $O(g^2T)$ 
(here we have $\mu \sim gT$), the gauge
condition (\ref{ggcond}) requires that $\tilde{G}^{>}$,
$\tilde{G}^{<}$ and $\tilde{G}^{C}$ be transversal.
If we further apply the transversality conditions in
the DSE one can verify that Eq.\ (\ref{com}) indeed holds.

Incorporating the transversality condition,
we can parametrize the
GF, $\tilde{G}^{>/</C}_{\a\b}$ and
$\d\tilde{G}^{>/</C}_{\a\b}$, 
in the following form:
\bee f_{\a\b}(X,{\mathbf q})
=T_{\a\b}f(X,{\mathbf q})\;,
\label{septrans} \eee
where $f$ stands for $\tilde{G}^{>/</C}$, $\tilde{G}^{(0)>/</C}$
or $\d\tilde{G}^{>/</C}$. We have also assumed that
$\delta\tilde{G}^{C}$ has the same structure as
$\tilde{G}^{(0)C}$, except that $\tilde{G}^{(0)C}$ is a colour
singlet. Note that we have separated all Lorentz indices into the
transversal projector $T_{\a\b}= g_{\a\b}-q_{\a}q_{\b}/q^2$ in
Eq.\ (\ref{septrans}). Then we can use Lorentz scalars $f$ to
express the Boltzmann equation (note that these $f$'s are different
from the one used earlier).

Under the assumption that
\bee \delta\tilde{G}^<_{\a\b}(X,{\mathbf q})
=\delta\tilde{G}^>_{\a\b}(X,{\mathbf q}) =\frac
12\delta\tilde{G}^{C}_{\a\b}(X,{\mathbf q}) \;,
\label{dg} \eee
and inserting Eqs.\ (\ref{septrans}) and (\ref{dg}) into
(\ref{trtiny}), we find
\bea
&&T_{\alpha\gamma} \Big[ q\cdot
\partial _X\delta \tilde{G}^C -gF_{\lambda\nu}q_{\nu}\partial
_{q\lambda}N^{(0)}\Big]
\nonumber\\
&&+g(F_{\alpha\beta}T_{\beta\gamma}
-T_{\alpha\beta}F_{\beta\gamma})\tilde{G}^{(0)C}
\nonumber\\
&&=i T_{\alpha\gamma}\Big\{ \tilde{G}^{(0)<}
\delta\tilde{\Pi}^{>}_{T}
-\tilde{G}^{(0)>}\delta\tilde{\Pi}^{<}_{T}\Big\}
\nonumber\\
&&-\frac 12 T_{\alpha\gamma}\Big[ i\tilde{\Pi}^{(0)<}_{T}
-i\tilde{\Pi}^{(0)>}_{T}\Big] \d\tilde{G}^{(0)C}\;,
\label{trtiny1}
\eea
where we used the following notation:
\bea
2 T_{\alpha\gamma} \delta\tilde{\Pi}^{>/<}_{T}&\equiv&
T_{\alpha\beta}\;\delta\tilde{\Pi}^{>/<}_{\beta\gamma}+
\delta\tilde{\Pi}^{>/<}_{\alpha\beta}\;\;
T_{\beta\gamma} \;, \nonumber\\
2 T_{\alpha\gamma} \tilde{\Pi}^{(0)>/<}_{T}&\equiv&
T_{\alpha\beta}\;\delta\tilde{\Pi}^{(0)>/<}_{\beta\gamma}+
\delta\tilde{\Pi}^{(0)>/<}_{\alpha\beta}\;\;T_{\beta\gamma} \;.
\eea
One can recognize in Eq.\ (\ref{trtiny1})
that the first term on the r.h.s.
is the collision and the second the
damping term, the damping rate being given by
$[i\tilde{\Pi}^{(0)<}_{T} -i\tilde{\Pi}^{(0)>}_{T}]$.
The physical interpretation and analysis of these terms
can be found in the recent papers by
Blaizot and Iancu \cite{blaizot,blaizot1}.

Equation (\ref{trtiny1}) is a linearized version of the
Boltzmann equation in a pure gluon plasma with respect to the
off-equilibrium fluctuations.
The linearized equation was previously derived in
Ref.\ \cite{blaizot}. 
In our approach, however, this equation was
obtained in the covariant background gauge, 
whereas in Ref.~\cite{blaizot} it was done in the
Coulomb background gauge. 
In the Coulomb gauge, the physical polarizations
are entirely contained in the spatial gluon propagator and are
independent of the Coulomb ghost. In the covariant gauge the
physical transverse degrees of freedom are mixed in all components
of the gluon propagator. Hence, the ghost diagrams are necessary
to cancel the unphysical polarization and to guarantee the
unitarity. Much as for the gluon,
one also needs to introduce covariant GFs for ghost fields and
formulate their evolution and transport equations.
Note that only the collision part contains
contributions from the ghost because they appear
in the SE diagrams.

\section{Summary and conclusions}

In this paper we have presented a systematic derivation of the
quantum Boltzmann equation for a pure gluon plasma. First we have
developed a functional method to derive the DSE in the BG-QCD.
The 1-PI vertex and the CGF were defined by the functional
derivatives of their generating functionals with respect to the
field average and the external source, respectively. The bare
vertex was derived by taking the functional derivative of the
classical action with respect to the corresponding field.

We have started our derivation by expressing the classical action
in the DeWitt notation, which, in our opinion, results in a simple
structure of the formalism. Then, taking a functional derivative of
the action with respect to the gluon field, we derived the
equation of motion. We recursively used the relations between the
generating functional for the CGF and that for the 1-PI vertex to
express a higher-rank GF in terms of the lower-rank CGFs and 1-PI
vertices. Using this method, we easily derived the DSE for the 2-
and 3-point GFs. The current approach has the great advantage that
it can treat a non-local and a local source term in the same way
and that it can produce all needed Feynman diagrams automatically. 
Hence our method is easy to implement by a computer algorithm that can
generate all Feynman diagrams for a given process.

We gave a heuristic discussion of the effects of the non-local
source kernel $K$ on the solution of the DSE for a free scalar
field. The role of $K$ is equivalent to that
of the initial density matrix. In the absence of a kernel, the
general solution of the DSE is the sum of the Feynman propagator,
which is the solution of the inhomogeneous DSE, and the solution
of the homogeneous DSE. The homogeneous solution involves a
particle momentum distribution function, which is just, 
in equilibrium, the Bose-Einstein distribution. 
We showed that, if the initial time is in the remote past, 
the effect of $K$ can be collected into 
the momentum distribution function. Thus, the
structure of the homogeneous solution is preserved.

The transport equation was derived from the DSE by performing the
gradient expansion. This expansion is justified only when the
quantum and kinetic scales are well separated.
We have introduced a mass parameter $\mu$ as a separating
point of these two scales. In ultra-relativistic heavy-ion
collisions, the weak coupling condition, $g\ll 1$, is most likely
to be fulfilled. In this case, 
low-momentum collective excitations that 
develop at the momentum scale $g\mu$ are well separated from 
typical hard gluon momenta $k\ge \mu$. We took the difference of
the two DSEs, which are in conjugate form, and then performed the
gradient expansion for the resulting equation under the above
conditions. Finally we used the gauge covariant GFs, which
are obtained by modifying the phase of the conventional GF
through Wilson links, to derive our final result of the Boltzmann
equation (\ref{treq1}). The sum of the two DSEs and its
subsequent gradient expansion gives the mass-shell constraint
equation for the gauge covariant GF.

The quantum kinetic equation was shown to be
a natural generalization of the classical one,
even though that it shows a more complicated non-Abelian structure.
A notable feature of our quantum result is that, as in the classical
case, it contains a term that corresponds to the colour
precession, the non-Abelian analogue to the Larmor precession for
particles with magnetic moments in a magnetic field. This
term is necessary to guarantee the gauge covariance of the quantum
kinetic equation.

The difference between the two conjugate DSEs (\ref{sch-dy-c1}) and
(\ref{sch-dy-c2}) gives the collision part. We have obtained this 
in a gauge covariant form and derived its
linearized form in the vicinity of equilibrium. Applying the
transversality requirement for the gauge covariant Wigner
functions, which arises from the background gauge condition in the
leading order, $O(g^2T)$, we have explicitly
identified the collision and the damping terms. A similar
equation was previously derived in Ref.~\cite{blaizot} in the
background Coulomb gauge. However, in the background covariant
gauge the results are more compact and have an explicit Lorentz
covariance. The contribution of the precession term in the kinetic
part of the Boltzmann equation was shown to be subleading, with
respect to the off-equilibrium fluctuations, thus it is not there
in the vicinity of equilibrium.

The current approach can be applied to study the propagation of
high energy partons (jets) through a hot and cold QCD medium.
This is because the coherent partons can be treated as the classical
background field. Over the past few years, substantial progress
has been made in understanding the induced gluon radiation in a
QCD medium \cite{mgxw,baier1}. However, owing to the complexity of the
problem, some idealized and simplified conditions were assumed in
order to obtain analytical or numerical results. In particular the
QCD medium is usually assumed to be in local chemical and thermal
equilibrium. We note that following the approach presented in
this work, one can study the energy loss of the fast parton and
the jet quenching via kinetic transport model suitable for
computer simulations. In this way one could study the influence of the
off-equilibrium effects on parton propagation and radiation.
We will address this problem in the future.

\begin{acknowledgments}

One of us, Q.W., acknowledges a fellowship from the Alexander von
Humboldt Foundation (AvH) and appreciated the help from D. Rischke.
This work is partially funded by DFG, BMBF and GSI. K.R.
acknowledges partial support from the Polish Committee for
Scientific Research (KBN-2P03B 03018). Stimulating comments and
discussions with R. Baier, J.-P. Blaizot, E. Iancu and S. Leupold
are acknowledged. Our special thanks go to X.-N. Wang for his
interest in this work and fruitful discussions, as well as to 
S. Mrowczy\`nski for interesting suggestions and comments. 
We acknowledge stimulating comments from H.-Th. Elze, his critical
reading of the manuscript and suggestions to include  
Eq.\ (\ref{mass}), which we finally added to the paper.

\end{acknowledgments}

\newpage

\appendix

\section{
Bare vertices}

\noindent In this appendix, we obtain bare vertices by taking
derivatives of the classical action with
respect to corresponding fields: 
\bea \Gamma _{mn}^{(0)}(Q^2)&=& \frac{\delta
^2S}{\delta Q^{\rho h}(x_1)
\delta Q^{\tau i}(x_2)}\nonumber\\
&=&\delta ^{hi}\int d^4u\delta ^4(x_1-u)
\Big[ g_{\rho\tau}\partial _u^2
-\partial _{u\tau}\partial _{u\rho}
+\frac{1}{\alpha}\partial _{u\tau}
\partial _{u\rho}\Big] \delta ^4(x_2-u)\;,
\eea
where $m=(\rho ,h,x_1)$ and $n=(\tau ,i,x_2)$; 

\bea
\Gamma _{mnp}^{(0)}(AQ^2)&=&
\frac{\delta ^3S}{\delta Q^{\rho h}(x_1)\delta A^{\tau i}(x_2)
\delta Q^{\eta j}(x_3)}\nonumber\\
&=&gf^{hij}\Big[ g_{\rho\eta}(\partial _3-\partial _1)_{\tau}
+g_{\rho\tau}
\Big( \partial _1-\partial _2+\frac{1}{\alpha}
\partial _3 \Big) _{\eta} \nonumber\\
&&+g_{\eta\tau}\Big( \partial _2-\partial _3-\frac{1}{\alpha}
\partial _1\Big) _{\rho} \Big] \int d^4u
\delta ^4(x_1-u)\delta ^4(x_2-u)\delta ^4(x_3-u)\;,
\eea
\bea
\Gamma _{mnp}^{(0)}(Q^3)&=&
\frac{\delta ^3S}{\delta Q^{\rho h}(x_1)\delta Q^{\tau i}(x_2)
\delta Q^{\eta j}(x_3)}\nonumber\\
&=&gf^{hij}\Big[ g_{\rho\eta}(\partial _3-\partial _1)_{\tau}
+g_{\rho\tau}(\partial _1-\partial _2)_{\eta}\nonumber\\
&&+g_{\eta\tau}(\partial _2-\partial _3)_{\rho} \Big] \int d^4u
\delta ^4(x_1-u)\delta ^4(x_2-u)\delta ^4(x_3-u)\;,
\eea
where $m=(\rho ,h,x_1)$, $n=(\tau ,i,x_2)$,  $p=(\eta ,j,x_3)$ and
$\partial _{1\rho}\equiv
\partial\delta (x_1-u)/\partial u^{\rho}$; 

\bea
\Gamma _{mnpq}^{(0)}(A^2Q^2)&=&
\frac{\delta ^4S}{\delta Q^{\mu a}(x_1)\delta Q^{\nu b}(x_2)
\delta A^{\lambda c}(x_3)\delta A^{\sigma d}(x_4)}\nonumber\\
&=&\Big[ -g^2f^{lda}f^{lcb}\Big( g_{\sigma\lambda}g_{\mu\nu}-
g_{\nu\sigma}g_{\mu\lambda}
+\frac{1}{\alpha}g_{\lambda\nu}g_{\mu\sigma}\Big)\nonumber\\
&&-g^2f^{lca}f^{ldb}\Big( g_{\sigma\lambda}g_{\mu\nu}-
g_{\mu\sigma}g_{\nu\lambda}
+\frac{1}{\alpha}g_{\sigma\nu}g_{\mu\lambda}\Big)\nonumber\\
&&-g^2f^{lcd}f^{lab}(g_{\mu\lambda}g_{\sigma\nu}-
g_{\mu\sigma}g_{\nu\lambda})\Big] \nonumber\\
&&\int d^4u
\delta ^4(x_1-u)\delta ^4(x_2-u)
\delta ^4(x_3-u)\delta ^4(x_4-u)\;,
\eea
\bea
\Gamma _{mnpq}^{(0)}(AQ^3)&=&
\frac{\delta ^4S}{\delta Q^{\mu a}(x_1)\delta Q^{\nu b}(x_2)
\delta Q^{\lambda c}(x_3)\delta A^{\sigma d}(x_4)}\nonumber\\
&=&\Big[ -g^2f^{lda}f^{lcb}(g_{\sigma\lambda}g_{\mu\nu}-
g_{\nu\sigma}g_{\mu\lambda})\nonumber\\
&&-g^2f^{lca}f^{ldb}(g_{\sigma\lambda}g_{\mu\nu}-
g_{\mu\sigma}g_{\nu\lambda})\nonumber\\
&&-g^2f^{lcd}f^{lab}(g_{\mu\lambda}g_{\sigma\nu}-
g_{\mu\sigma}g_{\nu\lambda})\Big]\nonumber\\
&&\int d^4u
\delta ^4(x_1-u)\delta ^4(x_2-u)
\delta ^4(x_3-u)\delta ^4(x_4-u)\;,
\eea
where $m=(\mu ,a,x_1)$, $n=(\nu ,b,x_2)$,
$p=(\lambda ,c,x_3)$ and $q=(\sigma ,d,x_4)$.
We can prove that $\Gamma _{mnpq}^{(0)}(Q^4)=
\Gamma _{mnpq}^{(0)}(AQ^3)$; 

\bee
\Gamma _{mn}^{(0)}(\ovl{C}C)=
\frac{\delta ^2S}{\delta C^{i}(x_2)\delta \ovl{C}^{h}(x_1)}
=\delta ^{hi}\int d^4u\delta ^4(x_1-u)
\partial _u^2\delta ^4(x_2-u)\;,
\eee
where $m=(h,x_1)$ and $n=(i,x_2)$; 

\bea
\Gamma _{mnp}^{(0)}(\ovl{C}CQ)&=&
\frac{\delta ^3S}
{\delta C^{i}(x_2)\delta \ovl{C}^{h}(x_1)
\delta Q^{\rho j}(x_3)}\nonumber\\
&=&gf^{hji}\int d^4u\delta ^4(x_1-u)\partial _{u\rho}
\delta ^4(x_2-u)\delta ^4(x_3-u) \;,
\eea
\bea
\Gamma _{mnp}^{(0)}(\ovl{C}CA)&=&
\frac{\delta ^3S}
{\delta C^{i}(x_2)\delta \ovl{C}^{h}(x_1)
\delta A^{\rho j}(x_3)}\nonumber\\
&=&gf^{hji}\int d^4u \Big\{ \delta ^4(x_1-u)
\partial _{u\rho}\Big[ \delta ^4(x_2-u)
\delta ^4(x_3-u)\Big] \nonumber\\
&&+\delta ^4(x_1-u)\delta ^4(x_3-u)
\partial _{u\rho}\delta ^4(x_2-u)\Big\}\;,
\eea
where $m=(h,x_1)$, $n=(i,x_2)$ and $p=(j,\rho ,x_3)$; 

\bea
\Gamma _{mnpq}^{(0)}(\ovl{C}CAQ)&=&
\frac{\delta ^4S}
{\delta C^{i}(x_2)\delta \ovl{C}^{h}(x_1)
\delta A^{\rho j}(x_3)\delta Q^{\eta k}(x_4)}\nonumber\\
&=&g^2f^{hja}f^{aki}g_{\rho\eta}
\int d^4u\delta ^4(x_1-u)\delta ^4(x_2-u)
\delta ^4(x_3-u)\delta ^4(x_4-u) \;,
\eea
where $m=(h,x_1)$, $n=(i,x_2)$,
$p=(j,\rho ,x_3)$ and $q=(k,\eta ,x_4)$.
\bea
\Gamma _{mnpq}^{(0)}(\ovl{C}CA^2)&=& \frac{\delta
^4S} {\delta C^{i}(x_2)\delta \ovl{C}^{h}(x_1)
\delta A^{\rho j}(x_3)\delta A^{\eta k}(x_4)}\nonumber\\
&=&g^2(f^{hja}f^{aki}+f^{hka}f^{aji})g_{\rho\eta}\nonumber\\
&&\cdot \int d^4u\delta ^4(x_1-u)\delta ^4(x_2-u)
\delta ^4(x_3-u)\delta ^4(x_4-u)\;,
\eea
with  $m=(h,x_1)$, $n=(i,x_2)$,
$p=(j,\rho ,x_3)$ and $q=(k,\eta ,x_4)$.

\vspace{2cm}

\section{
Relations and identities for Green functions}

\noindent Relations between the GF and the CGF
for the gluon field $Q$ are given by
\bea
\langle Q_{p}Q_{q}\rangle &=&\frac{\delta
^2W}{i\delta J_p\delta J_q} +\frac{\delta W}{\delta
J_p}\frac{\delta W}{\delta J_q}=
G_{pq}(Q^2)+\langle Q_p\rangle\langle Q_q\rangle \;,\nonumber \\
\langle Q_{m}Q_{n}Q_{p}\rangle
&=&\frac{\delta ^3W}{i^2\delta J_m\delta J_n\delta J_p}
+\frac{\delta ^2W}{i\delta J_m\delta J_n}\frac{\delta W}{\delta J_p}
+\frac{\delta ^2W}{i\delta J_n\delta J_p}\frac{\delta W}{\delta J_m}
+\frac{\delta ^2W}{i\delta J_m\delta J_p}\frac{\delta W}{\delta J_n}
\nonumber \\
&=&G_{mnp}(Q^3)+G_{mn}(Q^2)\langle Q_{p}\rangle
+G_{np}(Q^2)\langle Q_{m}\rangle
+G_{mp}(Q^2)\langle Q_{n}\rangle \;,\nonumber \\
\langle Q_{m}Q_{n}Q_{p}Q_{q}\rangle
&=&\frac{\delta ^4W}{i^3\delta J_m\delta J_n\delta J_p\delta J_q}
+\frac{\delta ^3W}{i^2\delta J_m\delta J_n\delta J_p}
\frac{\delta W}{\delta J_q}\nonumber \\
&&+\frac{\delta ^3W}{i^2\delta J_m\delta J_n\delta J_q}
\frac{\delta W}{\delta J_p}
+\frac{\delta ^3W}{i^2\delta J_m\delta J_p\delta J_p}
\frac{\delta W}{\delta J_n}\nonumber \\
&&+\frac{\delta ^3W}{i^2\delta J_n\delta J_p\delta J_q}
\frac{\delta W}{\delta J_m}
+\frac{\delta ^2W}{i\delta J_m\delta J_n}
\frac{\delta ^2W}{i\delta J_p\delta J_q}\nonumber \\
&&+\frac{\delta ^2W}{i\delta J_m\delta J_p}
\frac{\delta ^2W}{i\delta J_n\delta J_q}
+\frac{\delta ^2W}{i\delta J_m\delta J_q}
\frac{\delta ^2W}{i\delta J_n\delta J_p}\nonumber \\
&=&G_{mnpq}(Q^4)+G_{mnp}(Q^3)\langle Q_{q}\rangle \nonumber \\
&&+G_{mnq}(Q^3)\langle Q_{p}\rangle
+G_{npq}(Q^3)\langle Q_{m}\rangle \nonumber \\
&&+G_{mpq}(Q^3)\langle Q_{n}\rangle
+G_{mn}(Q^2)G_{pq}(Q^2)\nonumber \\
&&+G_{mp}(Q^2)G_{nq}(Q^2)+G_{mq}(Q^2)G_{np}(Q^2) \;,
\label{q-col}
\eea
where $G_{mn}(Q^2)$, $G_{mnp}(Q^3)$ and $G_{mnpq}(Q^4)$ are
the 2-, 3- and 4-point CGFs for $Q$. The above
equations are easy to prove if we recall, for example, that
$\langle Q_{p}Q_{q}\rangle
=Z^{-1}\delta ^2Z/(i^2\delta J_p\delta J_q)$
and $W=-i\ln Z$. For convenience, we sometimes use a
short hand notation $(mn)\equiv G_{mn}(Q^2)$,
$(mnp)\equiv G_{mnp}(Q^3)$, etc.

The relations between the GF and
the CGF for $Q$ and $\ovl{C}$/$C$ are:
\bea \langle \ovl{C}_{m}C_{n}\rangle &=&
\frac{\delta ^2W}{i\delta \ovl{\xi}_{n}\delta \xi _m}
=G_{mn}(\ovl{C}C)\;,\nonumber \\
\langle \ovl{C}_{m}C_{n}Q_{p}\rangle
&=&\frac{\delta ^3W}{i^2\delta \ovl{\xi}_{n}
\delta \xi _m\delta J_p}
+\frac{\delta ^2W}{i\delta \ovl{\xi}_{n}\delta \xi _m}
\frac{\delta W}{\delta J_p}\nonumber \\
&=&G_{mnp}(\ovl{C}CQ)+G_{mn}(\ovl{C}C)\langle Q_{p}\rangle
\;,\nonumber \\
\langle \ovl{C}_{m}C_{n}Q_{p}Q_{q}\rangle
&=&\frac{\delta ^4W}
{i^3\delta \ovl{\xi}_{n}\delta \xi _m\delta J_p\delta J_q}
+\frac{\delta ^3W}{i^2\delta \ovl{\xi}_{n}\delta \xi _m\delta J_p}
\frac{\delta W}{\delta J_q}\nonumber \\
&&+\frac{\delta ^3W}{i^2\delta \ovl{\xi}_{n}\delta \xi _m\delta J_q}
\frac{\delta W}{\delta J_p}
+\frac{\delta ^2W}{i\delta \ovl{\xi}_{n}\delta \xi _m}
\frac{\delta ^2W}{i\delta J_p\delta J_q}\nonumber \\
&=&G_{mnpq}(\ovl{C}CQ^2)+G_{mnp}(\ovl{C}CQ)\langle Q_{q}\rangle
\nonumber \\
&&+G_{mnq}(\ovl{C}CQ)\langle Q_{p}\rangle
+G_{mn}(\ovl{C}C)G_{pq}(Q^2)\;, \label{qc-col}
\eea
where $G_{mn}(\ovl{C}C)$, $G_{mnp}(\ovl{C}CQ)$ and
$G_{mnpq}(\ovl{C}CQ^2)$ are the 2-, 3- and 4-point CGFs
for $Q$ and $\ovl{C}$/$C$. We also use the following simplified
notation: $([mn])\equiv G_{mn}(\ovl{C}C)$, $([mn]p)\equiv
G_{mnp}(\ovl{C}CQ)$ and
$([mn]pq)\equiv G_{mnpq}(\ovl{C}CQ^2)$ etc.

A higher-rank CGF is related to lower-rank ones and 1-PI
vertices through the following identities
\bea
(mnp)&=&i\Gamma _{m'n'p'}(Q^3)(m'm)(n'n)(p'p)\;,\nonumber\\
(mnpq)&=&i\Gamma _{m'n'p'q'}(Q^4)(m'm)(n'n)(p'p)(q'q)\nonumber\\
&&+i\Gamma _{m'n'p'}(Q^3)(m'mq)(n'n)(p'p)\nonumber\\
&&+i\Gamma _{m'n'p'}(Q^3)(m'm)(n'nq)(p'p)\nonumber\\
&&+i\Gamma _{m'n'p'}(Q^3)(m'm)(n'n)(p'pq)\;,
\label{cgq1}
\eea
\bea
([mn]p)&=&i\Gamma _{n'm'p'}(\ovl{C}CQ)([mm'])([n'n])(p'p)
\;,\nonumber\\
([mn]pq)&=&i\Gamma _{n'm'p'q'}(\ovl{C}CQ^2)
([mm'])([n'n])(p'p)(q'q)\nonumber\\
&&+i\Gamma _{n'm'p'}(\ovl{C}CQ)
([mm']q)(n'n)(p'p)\nonumber\\
&&+i\Gamma _{n'm'p'}(\ovl{C}CQ)
([mm'])([n'n]q)(p'p)\nonumber\\
&&+i\Gamma _{n'm'p'}(\ovl{C}CQ) ([mm'])([n'n])(p'pq)\;,
\label{cgq2} \eea
which can easily be derived by repeatedly taking
the derivative with respect to $\langle Q\rangle$ for
$\delta \Gamma /\delta \langle Q\rangle=-J$.

The 4-point GF can be expressed in terms of
the lower-rank CGFs as follows:
\bea
\langle Q_{n'}Q_{p'}Q_{n}Q_{p}\rangle &=&
(n'p')(np)+(n'n)(p'p)+(n'p)(p'n)+(n'p'np)\nonumber\\
&=&(n'p')(np)+(n'n)(p'p)+(n'p)(p'n)\nonumber\\
&&+i\Gamma _{r's't'u'}(Q^4)(r'n')(s'p')(t'n)(u'p)\nonumber\\
&&+i\Gamma _{r's't'}(Q^3)(r'n'p)(s'p')(t'n)\nonumber\\
&&+i\Gamma _{r's't'}(Q^3)(r'n')(s'p'p)(t'n)\nonumber\\
&&+i\Gamma _{r's't'}(Q^3)(r'n')(s'p')(t'np) \;. \label{gf4q} \eea
Note that in the second equality the fifth
and sixth terms are equal. We can easily
identify in $\langle Q_{n'}Q_{p'}Q_{n}Q_{p}\rangle $ the
disconnected Green function $(n'p')(np)$. After dropping it, we
get the connected part $\langle Q_{n'}Q_{p'}Q_{n}Q_{p}\rangle _c$.

The 5-point GF can be expressed in terms of
the lower-rank CGFs as follows:
\bea
\frac 16 \Gamma _{mn'p'q'}^{(0)}(Q^4)
\langle Q_{n'}Q_{p'}Q_{q'}Q_{n}Q_{p}\rangle
&=&\frac 16 \Gamma _{mn'p'q'}^{(0)}(Q^4)
\Big[ (n'p'q'np)+(n'p')(q'np)\nonumber\\
&&+(n'q')(p'np)+(p'q')(n'np)+(n'n)(p'q'p)\nonumber\\
&&+(p'n)(n'q'p)+(q'n)(n'p'p)+(n'p)(p'q'n)\nonumber\\
&&+(p'p)(n'q'n)+(q'p)(n'p'n)+(np)(n'p'q')\Big] \;,
\label{gf5q1}
\eea
where the last term is a disconnected one. There are three
groups of terms that are equal, respectively: 
(2nd, 3rd, 4th), (5th, 6th, 7th) and (8th, 9th, 10th).
After collecting these terms, we get
\bea \frac 16 \Gamma _{mn'p'q'}^{(0)}(Q^4) \langle
Q_{n'}Q_{p'}Q_{q'}Q_{n}Q_{p}\rangle _c &=&\frac 16 \Gamma
_{mn'p'q'}^{(0)}(Q^4)
\Big[ 3(n'p')(q'np)+3(n'n)(p'q'p)\nonumber\\
&&+3(n'p)(p'q'n)+(n'p'q'np)\Big] \;,
\label{gf5q2}
\eea
where the 5-point CGF can be expanded as
\bea
(n'p'q'np)&=&
i\Gamma _{r's't'u'v'}(Q^5)(r'n')(s'p')(t'q')(u'n)(v'p)
\nonumber \\
&&+3i\Gamma _{r's't'u'}(Q^4)(r'n'p)(s'p')(t'q')(u'n)\nonumber \\
&&+3i\Gamma _{r's't'u'}(Q^4)(r'n'n)(s'p')(t'q')(u'p)\nonumber \\
&&+i\Gamma _{r's't'u'}(Q^4)(r'n')(s'p')(t'q')(u'np)\nonumber \\
&&+6i\Gamma _{r's't'}(Q^3)(r'n'n)(s'p'p)(t'q')\nonumber \\
&&+3i\Gamma _{r's't'}(Q^4)(r'n'np)(s'p')(t'q')\;.
\label{gf5q3}
\eea

The 4-point GF $\langle \ovl{C}_{m'}C_{n'}Q_{n}Q_{p}\rangle $ can
be expressed in terms of a lower-rank CGFs as follows:
\bea
&&\Gamma _{m'n'm}^{(0)}(\ovl{C}CQ)
\langle \ovl{C}_{m'}C_{n'}Q_{n}Q_{p}\rangle \nonumber \\
&&=\Gamma _{m'n'm}^{(0)}(\ovl{C}CQ)\Big\{([m'n']np)+([m'n'])(np)\Big\}
\nonumber \\
&&=\Gamma _{m'n'm}^{(0)}(\ovl{C}CQ)
\Big\{i\Gamma _{s'r't'u'}(\ovl{C}CQ^2)([m'r'])([s'n'])(t'n)(n'p)
\nonumber \\
&&+i\Gamma _{s'r't'}(\ovl{C}CQ)([m'r']p)([s'n'])(t'n)
+i\Gamma _{s'r't'}(\ovl{C}CQ)([m'r'])([s'n']p)(t'n)\nonumber \\
&&+i\Gamma _{s'r't'}(\ovl{C}CQ)([m'r'])([s'n'])(t'np)
+([m'n'])(np)\Big\} \;,
\label{gf4c}
\eea
where the last term is the disconnected one.

The 5-point GF
$\langle \ovl{C}_{m'}C_{n'}Q_{p'}Q_{n}Q_{p}\rangle $
can be written as
\bea
&&\Gamma _{m'n'p'm}^{(0)}(\ovl{C}CQ^2)
\langle \ovl{C}_{m'}C_{n'}Q_{p'}Q_{n}Q_{p}\rangle \nonumber\\
&&=\Gamma _{m'n'p'm}^{(0)}(\ovl{C}CQ^2)
\Big\{ ([m'n']n)(p'p)\nonumber\\
&&+([m'n']p)(p'n)+([m'n']p')(np)
+([m'n']p'np)+([m'n'])(p'np)\Big\} \;,
\label{gf5c}
\eea
where the fourth term is the disconnected one,
and the 5-point CGF
$G_{m'n'p'np}(\ovl{C}CQ^3)$ is given by:
\bea
&&([m'n']p'np)=i\Gamma _{[s'r']t'u'v'}
([m'r'])([s'n'])(t'p')(u'n)(v'p)\nonumber\\
&&+i\Gamma _{[s'r']t'u'}([m'r']p)([s'n'])(t'p')(u'n)
+i\Gamma _{[s'r']t'u'}([m'r'])([s'n']p)(t'p')(u'n)\nonumber\\
&&+i\Gamma _{[s'r']t'u'}([m'r'])([s'n'])(t'p'p)(u'n)
+i\Gamma _{[s'r']t'u'}([m'r'])([s'n'])(t'p')(u'np)\nonumber\\
&&+i\Gamma _{[s'r']t'u'}([m'r']n)([s'n'])(t'p')(u'p)
+i\Gamma _{[s'r']t'u'}([m'r'])([s'n']n)(t'p')(u'p)\nonumber\\
&&+i\Gamma _{[s'r']t'u'}([m'r'])([s'n'])(t'p'n)(u'p)\nonumber\\
&&+i\Gamma _{[s'r']t'}([m'r']np)([s'n'])(t'p')
+i\Gamma _{[s'r']t'}([m'r']n)([s'n']p)(t'p')\nonumber\\
&&+i\Gamma _{[s'r']t'}([m'r']n)([s'n'])(t'p'p)
+i\Gamma _{[s'r']t'}([m'r']p)([s'n']n)(t'p')\nonumber\\
&&+i\Gamma _{[s'r']t'}([m'r'])([s'n']np)(t'p')
+i\Gamma _{[s'r']t'}([m'r'])([s'n']n)(t'p'p)\nonumber\\
&&+i\Gamma _{[s'r']t'}([m'r']p)([s'n'])(t'p'n)
+i\Gamma _{[s'r']t'}([m'r'])([s'n']p)(t'p'n)\nonumber\\
&&+i\Gamma _{[s'r']t'}([m'r'])([s'n'])(t'p'np) \;,
\eea
where we use the following short-hand notations:
\bea
\Gamma _{[s'r']t'u'v'}&\equiv &\Gamma _{s'r't'u'v'}
(\ovl{C}CQ^3)\;,\nonumber\\
\Gamma _{[s'r']t'u'}&\equiv &\Gamma _{s'r't'u'}(\ovl{C}CQ^2)
\;, \nonumber\\
\Gamma _{[s'r']t'}&\equiv &\Gamma _{s'r't'}(\ovl{C}CQ)\;.
\eea

\newpage

\begin{figure}
\includegraphics{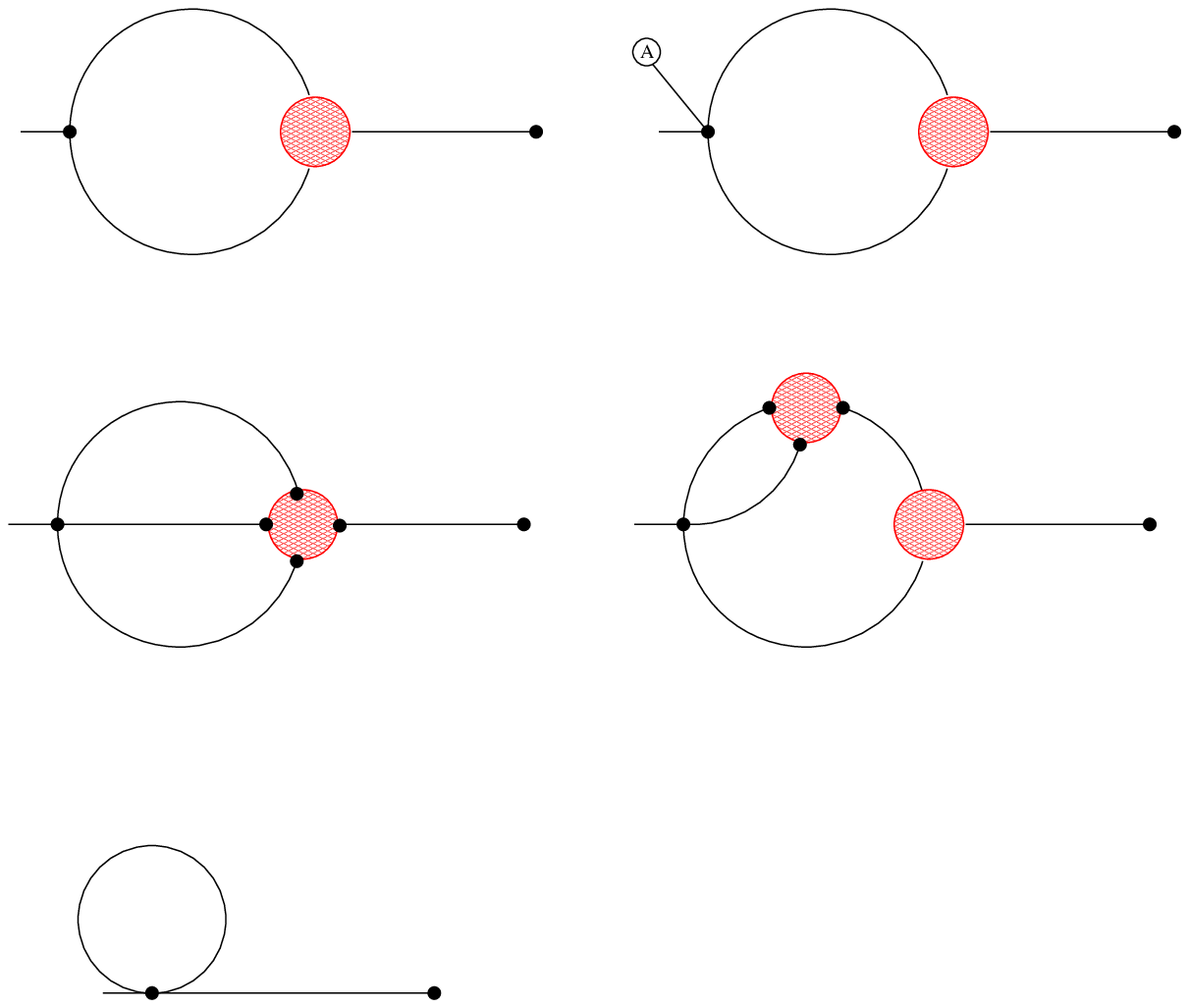}
\caption{Gluon loop contributions to $\Pi _{mn}G_{np}$,
corresponding to \eqrf{pi1}.}
\vspace*{.4in}
\end{figure}

\begin{figure}
\includegraphics{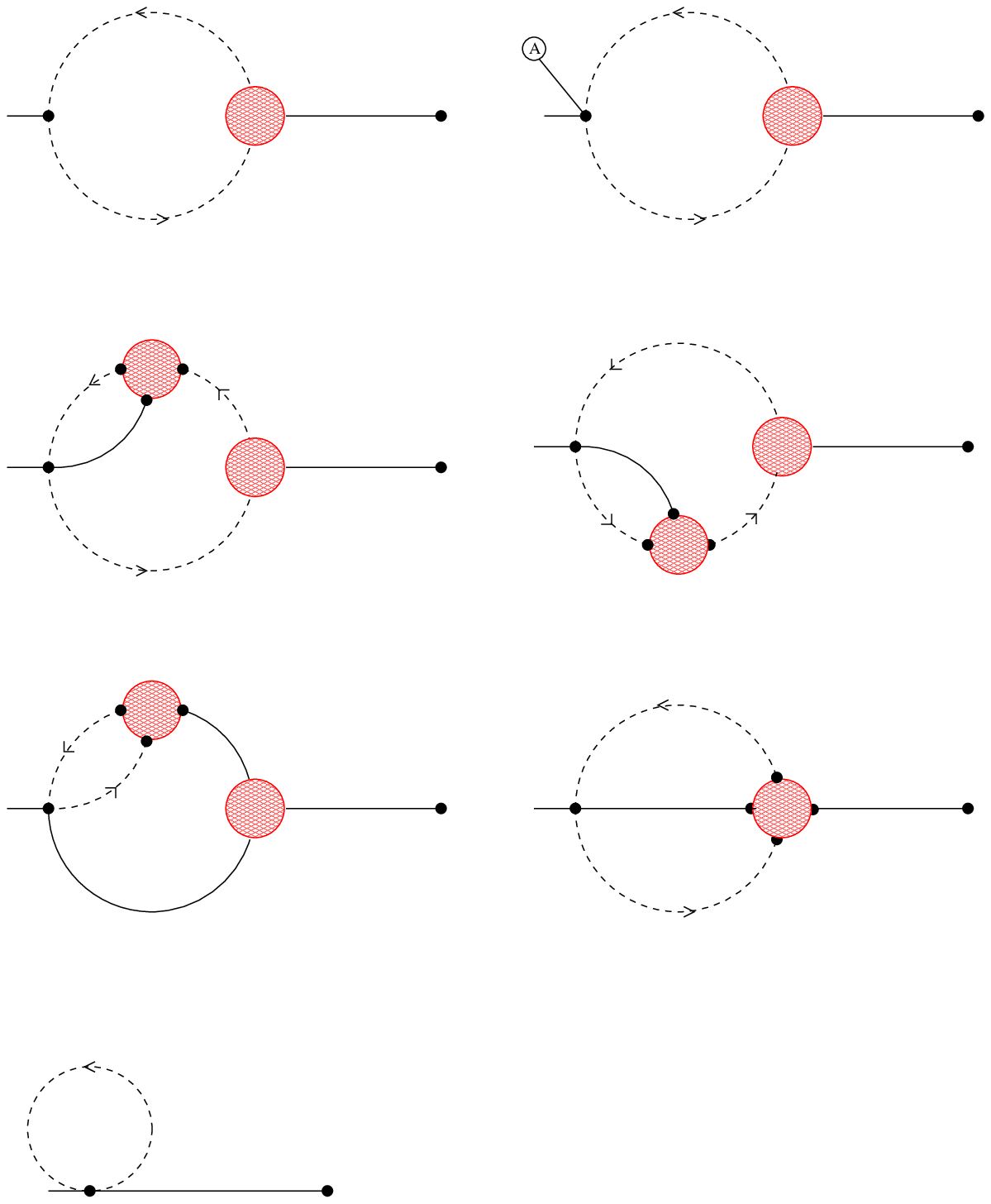}
\caption{Ghost loop contributions to $\Pi _{mn}G_{np}$,
corresponding to \eqrf{pi2}.}
\vspace*{.4in}
\end{figure}

\begin{figure}
\includegraphics{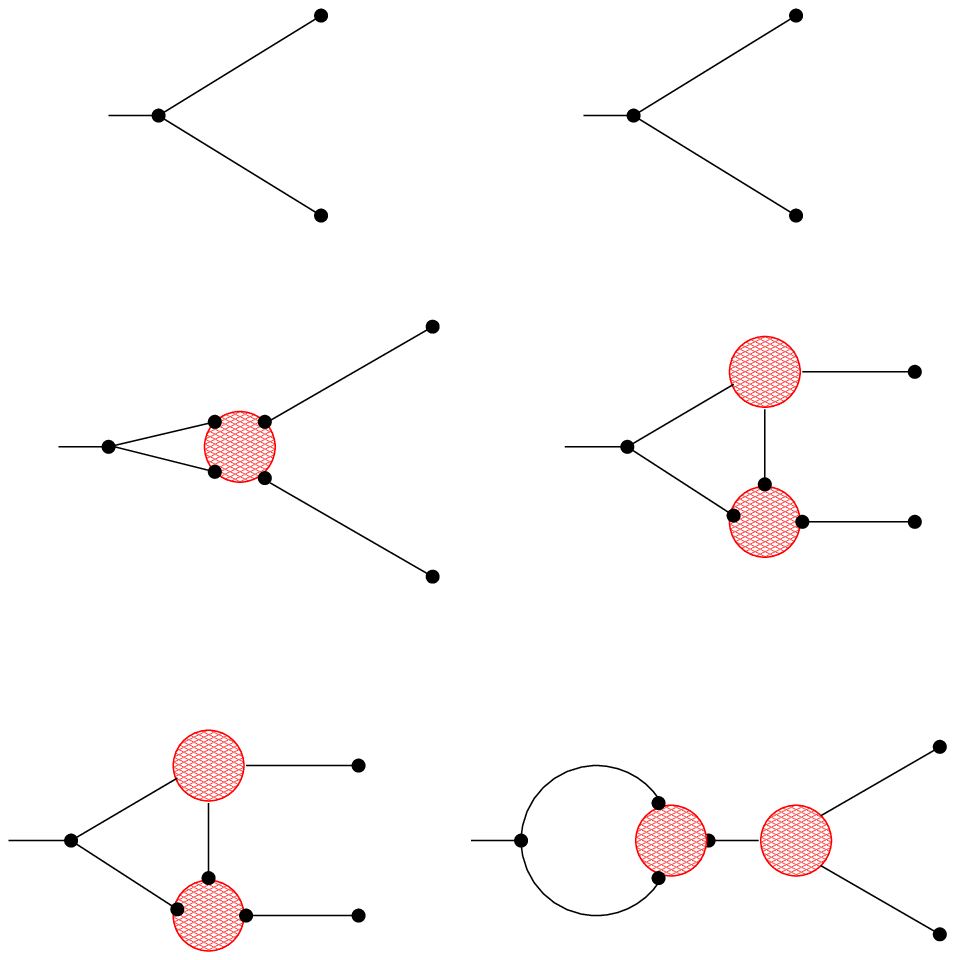}
\caption{Connected part of 4-point GF
$\langle Q_1Q_2Q_3Q_4\rangle $, corresponding to
\eqrf{gf4q}.}
\vspace*{.4in}
\end{figure}

\begin{figure}
\includegraphics{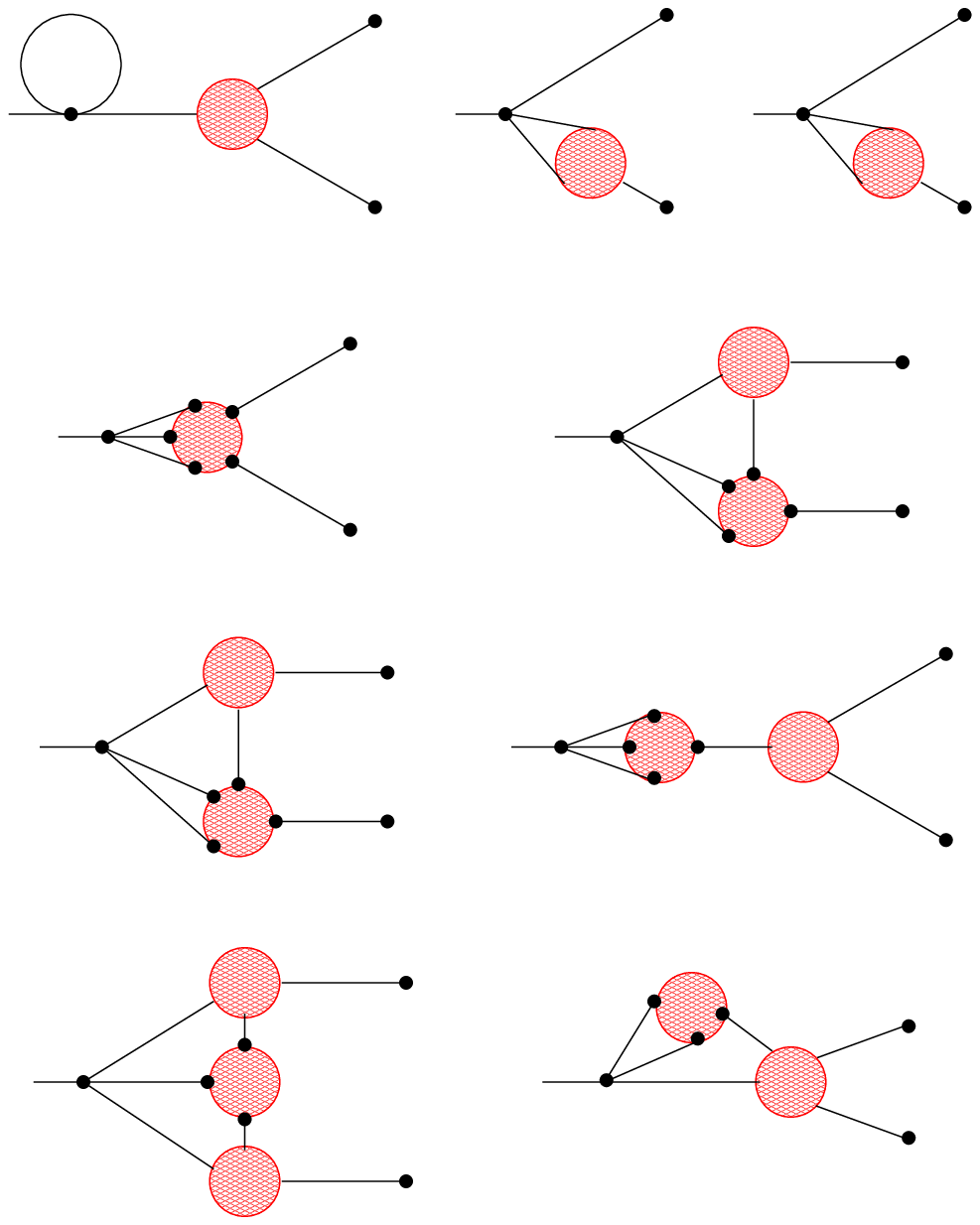}
\caption{Connected part of 5-point GF
$\langle Q_1Q_2Q_3Q_4Q_5\rangle $, corresponding to
\eqrf{gf5q2}. }
\vspace*{.4in}
\end{figure}

\begin{figure}
\includegraphics{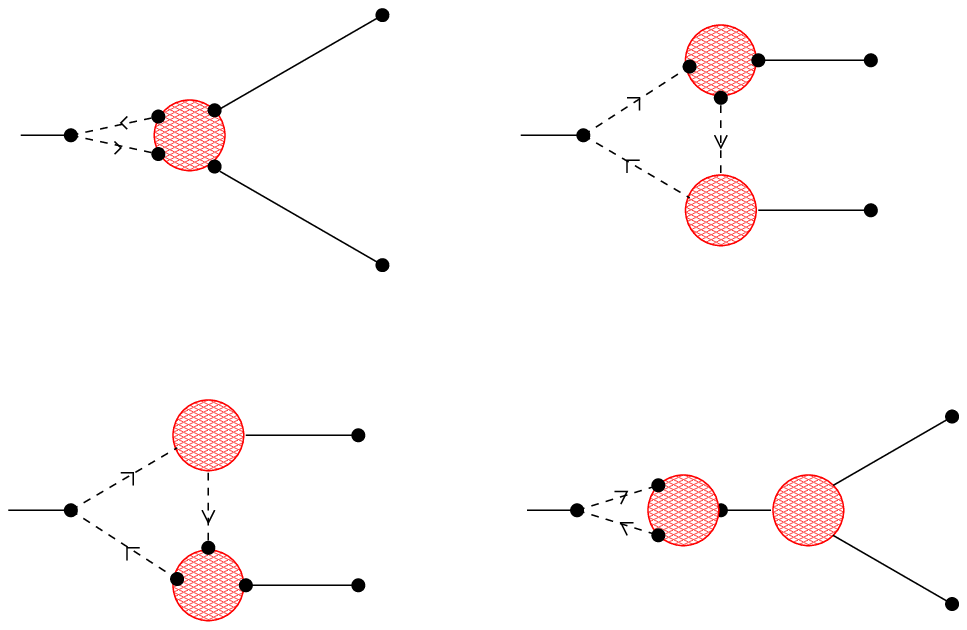}
\caption{Connected part of 4-point GF
$\langle \ovl{C}CQ_1Q_2\rangle $, corresponding to
\eqrf{gf4c}. }
\vspace*{.4in}
\end{figure}

\begin{figure}
\includegraphics{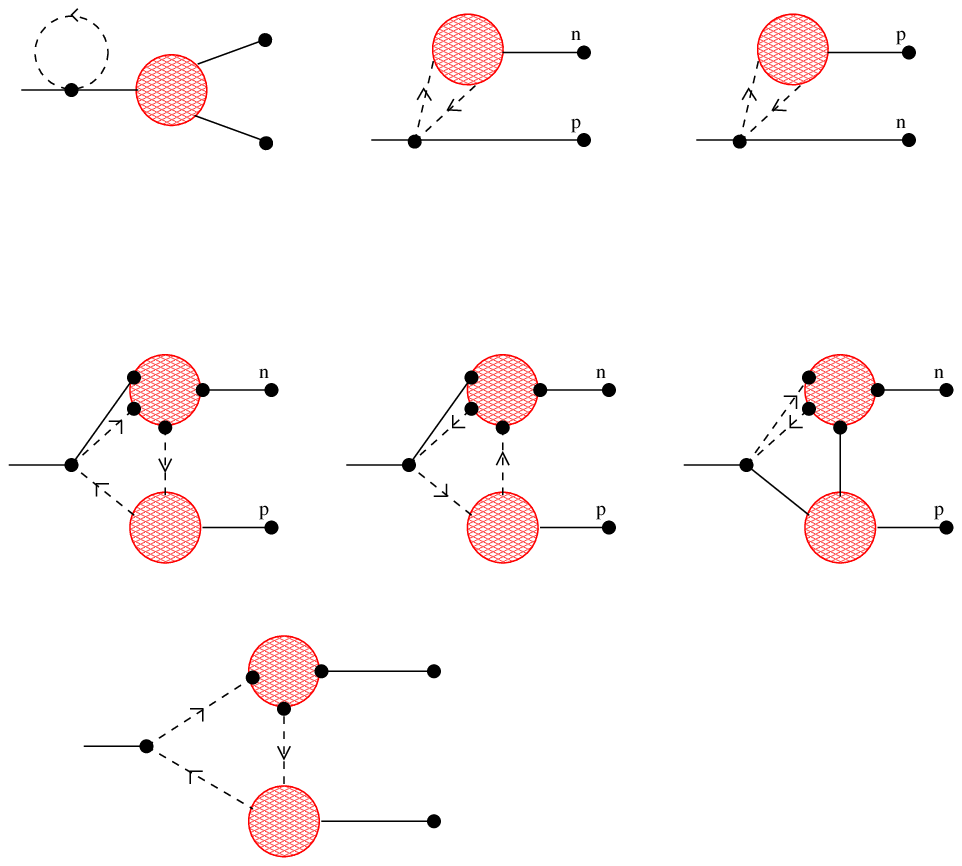}
\caption{Connected part of 5-point GF
$\langle \ovl{C}CQ_1Q_2Q_3\rangle $, corresponding to
\eqrf{gf5c}. }
\vspace*{.4in}
\end{figure}

\end{document}